\documentclass[%
aip,
amsmath,amssymb,
reprint,%
]{revtex4-2}

\usepackage{graphicx}
\usepackage{dcolumn}
\usepackage{bm}

\usepackage[utf8]{inputenc}
\usepackage[T1]{fontenc}
\usepackage{mathptmx}
\usepackage{etoolbox}
\usepackage{xcolor}
\usepackage{chngcntr}

\makeatletter
\def\@email#1#2{%
	\endgroup
	\patchcmd{\titleblock@produce}
	{\frontmatter@RRAPformat}
	{\frontmatter@RRAPformat{\produce@RRAP{*#1\href{mailto:#2}{#2}}}\frontmatter@RRAPformat}
	{}{}
}%
\makeatother
\begin{document}
	
	\preprint{AIP/123-QED}
	
	\title{Surfactant Induced Catastrophic Collapse of Carbon Black Suspension used in Flow Battery Application}
	\author{KangJin Lee}
	\author{Mohan Das*}
	\email{fxm212@case.edu}

	\author{Matthew Pitell}
	\author{Christopher L. Wirth}
	\affiliation{%
	Department of Chemical and Biomolecular Engineering, Case Western Reserve University, 10900 Euclid Ave, Cleveland, USA, 44106.
	}%

	
	\date{\today}
	
	\begin{abstract}
		 Carbon black particles act as electronically conductive additives in the slurry electrodes used in electrochemical redox flow batteries. Stability and dispersion of the carbon black particles in a slurry electrode are critical parameters for its storage and the efficient functioning of the battery. Modifying the carbon black slurry formulation with the addition of a nonionic surfactant could potentially impart desired properties such as good particle dispersion, gravitational stability and flowability imparting better performance of the flow battery. Matching the typical slurry electrode formulation, we dispersed carbon black particles in 1 M H$_2$SO$_4$ with volume fraction $\phi$ = 0.01 to 0.06 and c$_{surf.}$ = 0, 0.05 and 0.1 M. Rheological investigation reveals that the carbon black suspensions behave like colloidal gels. Sedimentation kinetics of the process was measured by tracking the height of the particle bed over time in a cuvette using a custom camera set-up. The sedimentation dynamics here clearly resembled that of a gel collapse. At short times we observe fast sedimentation associated with structural collapse of the gel and at long times very slow sedimentation associated with compaction of the sediment. Addition of a nonionic surfactant (Triton X-100) to the solvent above CMC at $\alpha$ (= \textit{c$_{surf.}$/c$_{CB}$)} $<$  0.7 improves particle dispersion and increases gel elasticity. However, for \textit{$\alpha >$} 0.7 leads to the formation of a weaker gel that exhibits 'catastrophic collapse' under gravity and has a lower viscosity. 
	\end{abstract}
	
	\maketitle
	
		%
	
	\section{\label{sec:level1}Introduction}
	
Electrochemical or `redox' flow batteries (RFB) are useful alternates for solid-state batteries in grid-scale electrical energy storage used for renewables.\cite{dunn2011electrical,park2016material,gurieff2019performance} RFBs store energy separately in external reservoirs and their power output is determined by the active surface area within the electrochemical cell stack which is decoupled from the reservoirs. This allows for their easy scale up depending on the end user demand without facing space constraints and fire hazards associated with solid-state batteries.\cite{park2016material} In RFBs, the redox reactions occur on electrode surface within the cell that provides long life cycles (> 10000 cycles, $\tilde{20}$ years)\cite{dunn2011electrical} and avoid problems such as lithium intercalation frequently encountered in the limited internal space of solid-state lithium-ion batteries.\cite{zhao2015chemistry}  
\par As an alternative to the traditional RFBs which present toxicity and cost challenges, an all-iron flow battery has been developed because of its use of single active element that is cheap, abundant, and environment friendly.\cite{hawthorne2014studies,petek2015slurry} When an all-iron flow battery is used in conventional RFB designed with stationary electrodes, the iron metal is plated on the electrode surface and stored in the electrochemical cell. This limits the amount of energy that can be stored and couples the energy and power capabilities.\cite{hawthorne2014studies} To resolve these issues, as in many RFBs, a flowable slurry electrode is used that contains electronically conducting particles dispersed in an ionically conducting electrolyte.\cite{duduta2011semi,petek2015slurry} The slurry is pumped through an electrochemical reactor allowing for charge and discharge by chemical reduction and oxidation (redox) reaction of the electroactive species. At sufficiently high volume fraction of the conducting particles an electrically conducting network is formed allowing the redox reaction to occur on the particle surface.\cite{backhurst1969preliminary,sabacky1977electrical,gao2007effective,dennison2014situ} Moreover, the slurry is pumped back into the reservoir thus removing the iron metal from the cell. Use of the slurry electrode not only resolves the iron plating problem but also provides advantages such as increased surface area for redox reactions.\cite{petek2015slurry}
\par Carbon black particles have shown their advantages in being used as a conductive additive for the electrode in an all-iron flow battery as they allow electrical percolation at relatively low volume fractions.\cite{kastening1997design,richards2016mixed} Additional advantages come from their structure, inherent chemical and mechanical resistance, abundance, and low cost.\cite{parant2017flowing,mourshed2021carbon} However, long term efficiency of slurry based RFBs presents challenges such as circulating a high viscosity slurry and particle sedimentation. As a result, for carbon black based slurry electrodes, it is essential to understand the flow behavior and stability of carbon black particles as they are circulated through electrochemical cell and stored in the reservoir. Carbon black particles when dispersed in aqueous mediums at low pH such as in the case of slurry electrodes used in all-iron flow batteries, tend to form aggregates.\cite{li2005polymer,xu2007particle,parant2017flowing} Particle aggregation negatively affects suspension stability by promoting sedimentation. However, in such systems at low enough particle concentration a space spanning network of particles is formed which can support stress, a process known as gelation.\cite{zaccarelli2007colloidal} However, gels are also known to exhibit sedimentation for a wide range of materials and interparticle attraction strength \textit{U}.\cite{allain1995aggregation,poon1999delayed,starrs2002collapse,manley2005gravitational,buscall2009towards,teece2011ageing,harich2016gravitational} Their sedimentation dynamics can be controlled by changing particle volume fraction and attraction strength.
\par Nonionic surfactants have been demonstrated to be good dispersants for carbon black particles.\cite{ma1992mixed,gupta2005adsorption} Moreover, they are known to improve performance of lithium-ion flow battery systems that use carbon black slurry.\cite{madec2015surfactant} When the surfactant concentration is high enough to form a monolayer on the surface of the carbon black particles, it enables a homogeneous dispersion of the particles, optimizing the electrochemical performance.\cite{porcher2010optimizing} While we expect the dispersion of particles as well as electrochemical performance to improve, the influence of surfactant on the gravitational stability of the suspension while in storage as well as its ease of flow are crucial parameters to be evaluated. It is well known that gravitational stability of a suspension can be improved either by using very small particles so that their Brownian motion will keep them suspended, matching the density of the dispersed particles with the dispersion medium, increasing the viscosity of the dispersion medium (to slow down particle sedimentation), or the formation of a gel that can slow down or arrest sedimentation completely.\cite{kim2007gravitational} Furthermore, improvements made in stability and flow of slurry electrodes can be utilized for other electrochemical applications such as flow capacitors, flow capacitive di-ionization cells, and electrochemical reactors (such as electrosynthesis of organic compounds).
\par In this paper, we report our study on gravitational stability and flow of carbon black suspensions at particle volume fractions matching that of slurry electrodes used in all-iron RFBs. We use activated carbon black particles dispersed in 1 M H$_2$SO$_4$ without and with the presence of a nonionic surfactant (Triton X-100). In the absence of surfactant, van der Waals attraction between carbon black particles leads the suspensions to behave like weak gels. Interestingly, the suspensions exhibit gel collapse under gravitational stress. Addition of a nonionic surfactant alters the interparticle attraction and influences the dynamics of gel collapse. We determine the dynamics of gel collapse using time-lapse video imaging where particle bed height is captured as a function of time. The gel collapse dynamics follows an exponential relaxation process. Moreover, in some cases the particle bed exhibits a very fast catastrophic collapse. We use rheology to determine the liquid-solid transition and yield stress of our suspensions to relate gel elasticity with gravitational stability. We show that a critical mass ratio of carbon black particles to that of surfactant is key to ensure desired suspension stability and flowability.
	
\section{\label{sec:level2}Materials and methods}
	
The critical micelle concentration (CMC) of the surfactant in dispersion medium was determined by measuring the surface tension of deionized water ($\rho$ = 18.2 M$\Omega$ cm) and 1 M  H$_2$SO$_4$ (96\%, Fisher Chemical) with different concentrations of Triton X-100 (M$_{w, avg}$: 625, Sigma-Aldrich), a non-ionic surfactant. Stock solution (165 mM) of Triton X-100 was first prepared in ultra-pure deionized water and a serial dilution was carried out to cover a large range of surfactant concentrations from 0.001-100 mM. Similar solutions of Triton X-100 were prepared with 1 M H$_2$SO$_4$. The surface tension of the solvents was measured via pendant drop method with a goniometer (Ossila contact Angle Goniometer). Moreover, dynamic light scattering (Zetasizer Nano SZ ZEN3600, Malvern) was used to measure the hydrodynamic radius (R$_H$) of the micelles above CMC in both solvents. 
	
\subsection{\label{sec:level2.1}Suspension preparation}
	
Carbon black suspension with composition similar to the RFB application was prepared by dispersing activated carbon black particles (YP-50, Kuraray) in 1 M H$_2$SO$_4$ with different surfactant concentrations c$_{surf.}$ = 0, 0.05 \& 0.1 M. 
Dry carbon black particles (YP-50, Kuraray) were added to the solvent in 20 mL glass vials followed by addition of the solvent. Particle concentrations were chosen based on the scope of interest in the flow battery application and six samples for each of the solvent with carbon black concentration of $\phi$ = 0.01 to 0.06 (corresponding to 0.02 to 0.12 g/ml) were prepared. The samples were stirred with a magnetic stirrer overnight ($>$ 8 h) to ensure consistency in the dispersion of particles in the solvent. Note that in order to isolate the role of surfactants on the properties of the carbon black suspension, we did not add any salt that is an important component of the slurry electrode formulation.

\subsection{\label{sec:level2.2}Characterization}
It is known that carbon black suspensions are extremely sensitive to their preparation method.\cite{akuzum2017effects} Hence, we measured the particle size distribution for dispersed carbon black particles in suspensions prepared using different methods. Carbon black suspension ($\phi$ = 0.01) was prepared in four different ways:  (a) stirring for $>$ 8 h using magnetic stirrer bar, (b) stirring followed by high shear homogenization for 10 min (T-25 Ultra Turrax homogenizer, IKA), (c) stirring followed by sonication for 30 min (CPX5800, Fisherbrand), and (d) a combination of all three methods. Microscopy images captured at 10$\times$ magnification (IXplore Standard, Olympus) were analyzed using ImageJ software for measuring particle size distribution. Based on our preliminary observations (see Figure S1, SI) we chose stirring process for sample preparation. Additionally, we captured the changes in particle surface morphology before and after dispersing in 1 M H$_2$SO$_4$ using a scanning electron microscope or SEM (Apreo 2S, ThermoFisher). For this the samples were deposited on a glass slide and dried in a hot air oven (Isotemp, FisherBrand) at 100 $^{\circ}$C for a week to ensure complete solvent evaporation. The samples were coated with platinum (Denton Desk IV Sputter Coater) prior to SEM imaging and the images were captured with 5 kV accelerating voltage.
	
\subsection{\label{sec:level2.3}Adsorption Isotherm Measurements}
To measure the adsorption of the nonionic surfactant (Triton X-100) on carbon black particles, suspensions with a fixed particle concentration (c$_{CB}$ = 6 g/ml or $\phi$ = 0.03) with varying concentration of surfactant (c$_{surf.}$) were prepared. Here $\alpha$ (=c$_{surf.}$/c$_{CB}$), was varied from 0.3-1.7. After mixing the samples overnight (> 8 h), they were allowed to sediment. The clear supernatant was collected and filtered (Fisherbrand - PTFE, 450 nm pore size) to remove any remaining carbon black particles. Concentration of the free surfactant in the supernatant (c$_{free}$) was measured using UV-vis spectrophotometer (Cary 3500, Agilent), in the range of wavelengths 200-500 nm with 1M H$_2$SO$_4$ as the baseline solvent.

\subsection{\label{sec:level2.4}Sedimentation study}
The sedimentation rate was measured by tracking the bed height of the carbon black particle suspension as it sediments with time. A custom optical set up comprising of a CMOS camera (CS235MU, Thorlabs) equipped with a zoom lens (6.5$\times$, Thorlabs), quartz cuvette (CV10Q35, Thorlabs) and a white light LED illumination source was used. Three sample cuvettes were used at the same time to ensure repeatability. The overnight stirred sample was filled in the cuvette while carefully avoiding bubble formation, covered with a plastic lid and sealed by wrapping parafilm to prevent any evaporation during the experiment. Sample cuvette was then placed between the camera and the light source to obtain images with a sharp contrast. Three sample cuvettes were used simultaneously for each suspension formulation. Images were recorded at 0.1, 0.25 and 1 fps depending on the total duration of the experiment. ImageJ and in-house developed MATLAB (MathWorks, Inc.) algorithms were used to analyze the 8-bit grayscale images to precisely locate the height of the particle bed at the solid-liquid interface represented by the bright (liquid phase) and dark (solid phase) regions. Additional details on image processing are available in SI. 

\begin{figure}[h]
	\centering
	\includegraphics[width=1\linewidth]{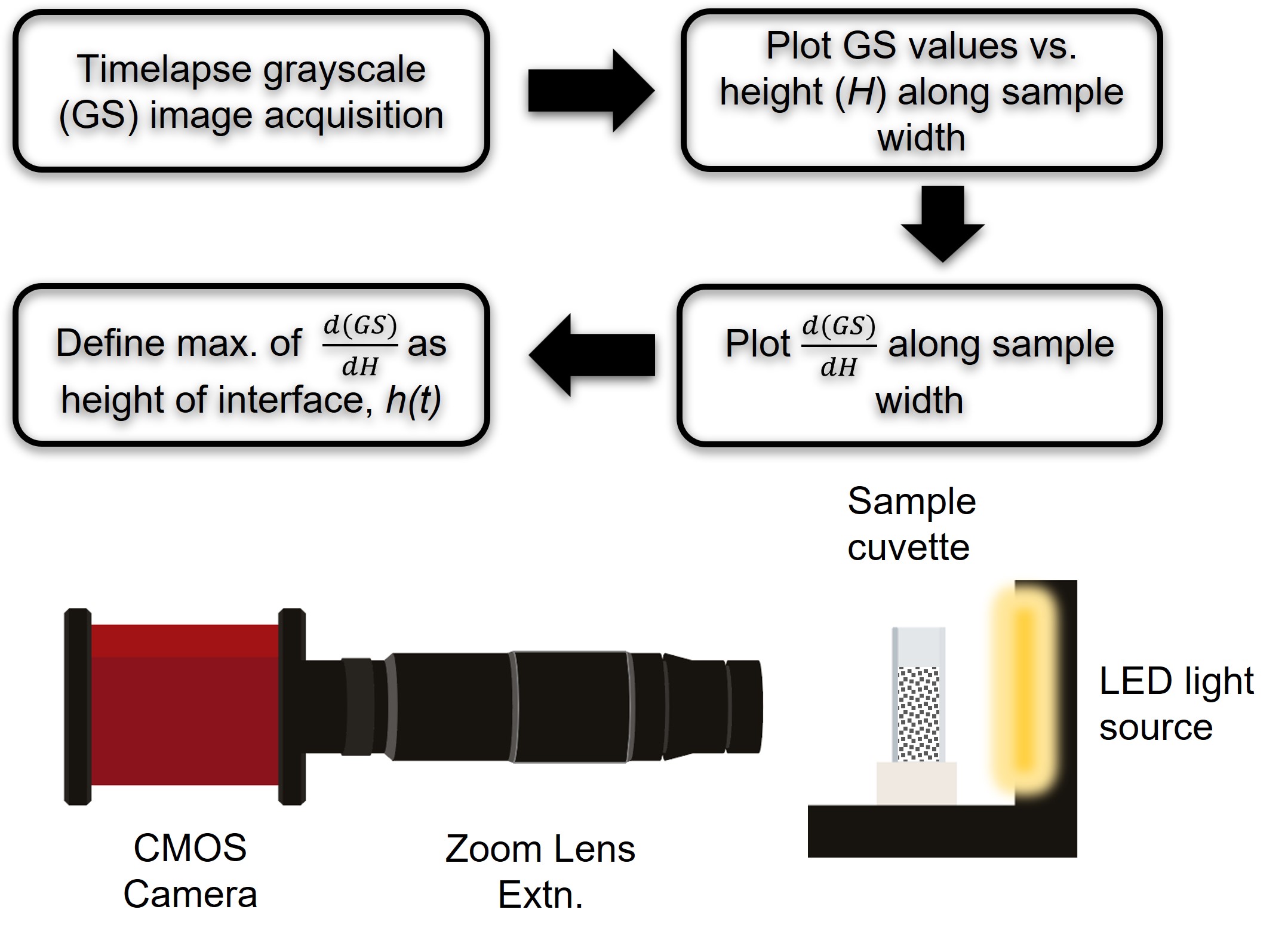}
	\caption[Fig1]{Schematic of the custom setup for capturing sedimentation kinetics.}
	\label{fig:fig1}
\end{figure}

\par The maximum packing ($\phi_{m}$) of the carbon black suspensions after sedimentation experiment was determined by removing the clear supernatant using a micropipette (100$\mu$l at a time) without disturbing the particle bed. Additionally, the maximum packing limit ($\phi_{max}$) of the carbon black suspension ($\phi_{m}$ = 0.05) in different solvents used here was determined by centrifugation of 50 ml sample vials at 7000$\times$g for 45 min (Sorvall ST 8, ThermoFisher). The clear supernatant was removed by decanting leaving behind a packed bed of carbon black particles. In both measurements, volume fraction of the carbon black in the sediment was calculated from mass of the sample before and after drying in a hot air oven at 100 $^{\circ}$C and after accounting for the mass of non-evaporated H$_2$SO$_4$.

\subsection{\label{sec:level2.5}Rheology}
Rheological characterisation of carbon black suspensions was carried out at 25 $^{\circ}$C using a strain-controlled rheometer (Haake Mars III, Thermofisher). Samples were loaded between a rough cone-plate geometry (cone angle: 1$^{\circ}$, dia: 35 mm) and a solvent trap containing a high absorption fabric soaked in water was used to maintain a humid environment around the sample. This arrangement helped in preventing solvent evaporation for the duration of the measurement. The sample was sheared at 1000 s$^{-1}$ for 200 s until steady stress values were obtained. This step also called rejuvenation eliminates any shear history associated with sample preparation and loading. Rejuvenation was followed by high to low shear rate ramp ($\dot{\gamma}$ = 1000-0.01 s$^{-1}$) until steady stress values were achieved. We followed the time evolution of sample viscoelasticity (or ageing) immediately after rejuvenation. This was done by performing an oscillatory shear measurement in the linear regime called dynamic time sweep (DTS) performed at $\gamma$ = 0.001 and $\omega$ = 1 rad s$^{-1}$ for waiting time t$_w$ = 4000 s. Following this step, the yield stress $\sigma_y$ (a maximum of elastic stress, G'$\gamma$)\cite{walls2003yield} was determined using dynamic strain sweep (DSS) measurement performed from $\gamma$ = 0.001 - 10 at $\omega$ = 1 rad s$^{-1}$.

	\section{\label{level3}Results and Discussion}
	\subsection{\label{level3.1}Particle size distribution}
\begin{figure}[h]
	\centering
	\includegraphics[width=1\linewidth]{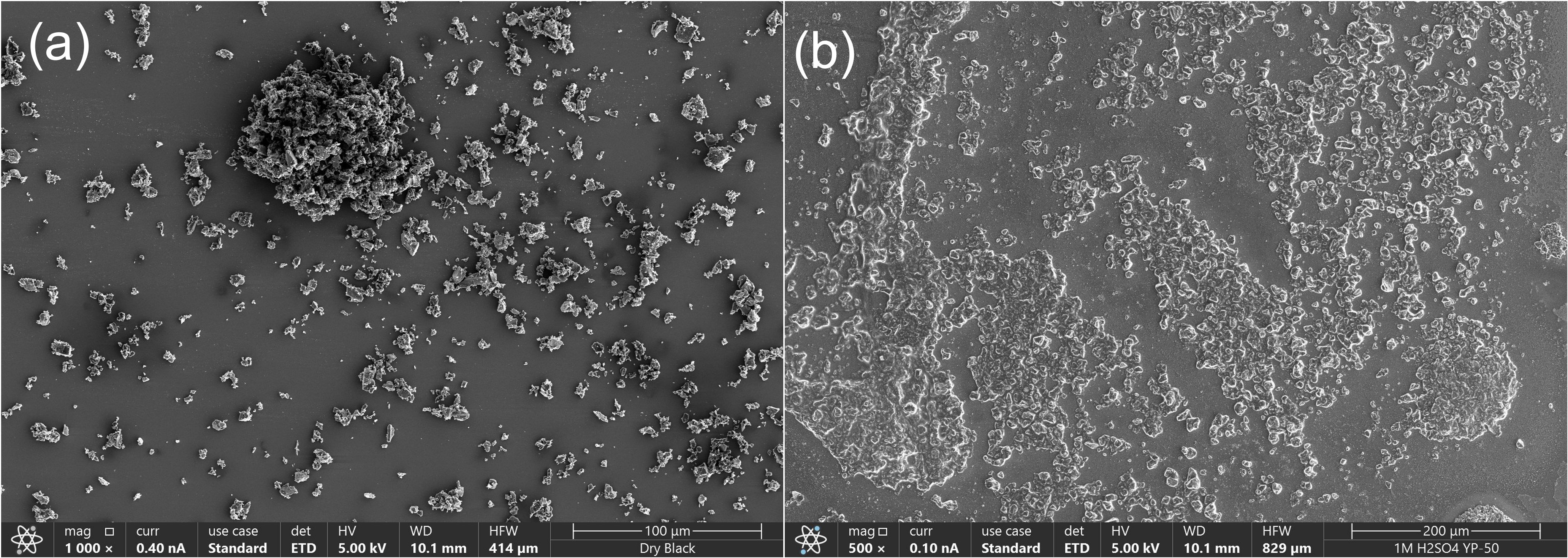}
	\caption[Fig2]{SEM images of YP-50 carbon black particles (a) before (b) after dispersing in 1 M H$_2$SO$_4$.}
	\label{fig:fig2}
\end{figure}

Carbon black particles used in our study are activated and hence are more porous and will have several oxygen containing functional groups such as hydroxyl groups, carboxyl groups, lactone groups on their surface.\cite{boehm1994some} Presence of such polar groups improves their wettability and helps in better dispersion in aqueous mediums compared to non-activated carbon blacks.\cite{bhatnagar2013overview} However, their high carbon content $>$ 95 \% in quasi-graphitic form makes them highly hydrophobic as well.\cite{antonucci1989influence,ridaoui2006effect} Figure \ref{fig:fig2} shows SEM images of carbon black particles in dry state and after dispersion in 1 M H$_2$SO$_4$ by stirring overnight. We observe very large particle aggregates in dry state with very rough surface morphology. Direct imaging suggested the aggregate size decreased after dispersion in 1 M H$_2$SO$_4$ (see Figure \ref{fig:fig2}). In Figure \ref{fig:fig2} (b) we observe the presence of a thin liquid film between smaller particle aggregates despite drying the sample in the oven at 100 $^{\circ}$C for more than 24 h. We attribute this non-dried layer to pure H$_2$SO$_4$ as the evaporation of water would increase the concentration of H$_2$SO$_4$ in the aqueous medium. Pure H$_2$SO$_4$ has a boiling point of around 337 $^{\circ}$C and would remain adsorbed to the particle surface after drying at 100 $^{\circ}$C.   

\par
\begin{figure}[h]
	\centering
	\includegraphics[width=0.75\linewidth]{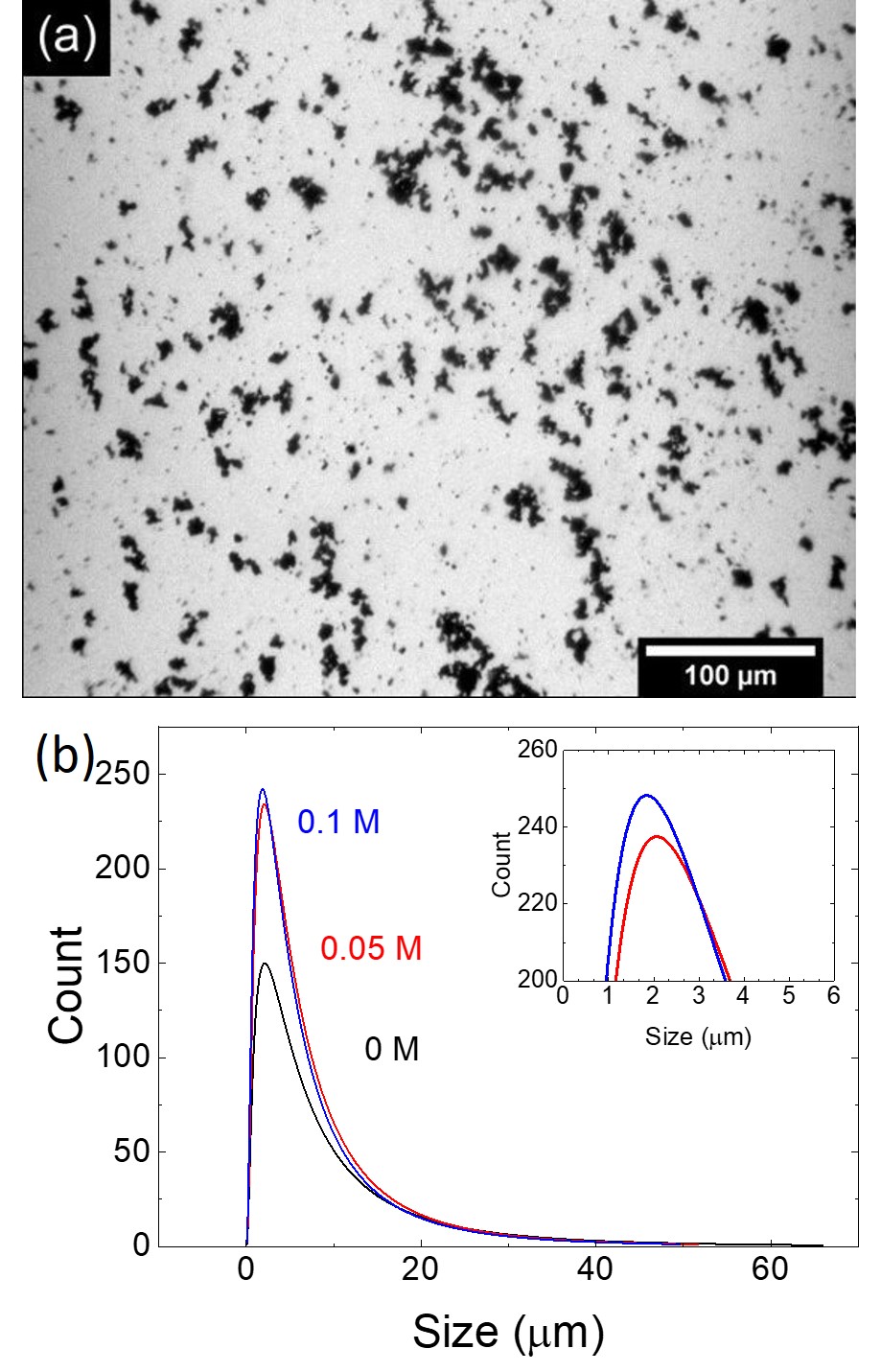}
	\caption[Fig3]{(a) Micrograph of carbon black suspension ($\phi$ = 0.01) captured at 20$\times$ magnification (b) Size distribution of carbon black particles after dispersing in 1 M H$_2$SO$_4$ at different surfactant concentrations (inset: difference in size distribution between different surfactant concentrations).}
	\label{fig:fig3}
\end{figure}

 Dispersion is crucial for long term stability of particle suspensions. It has been reported previously that the dispersion of carbon black particles is dependent on various parameters such as dispersion time, dispersion protocol, pH and ionic strength of the dispersion medium and presence of surface functional groups.\cite{medalia1964dispersant,akuzum2017effects} Moreover, addition of other molecules such as surfactants can further enhance or disrupt the carbon black dispersion.\cite{sis2009effect,madec2015surfactant,subramanian2021aqueous} We observed that mechanical stirring leads to the best dispersion of carbon black particles. Particle size distribution was not affected by changing the stirring time beyond 1 h and all the samples were stirred for over 12 h before further measurements. Addition of surfactant leads to a slight increase in the particle size distribution as shown in Figure \ref{fig:fig3} (b) with the particle count peak shifting towards values lower than 2.5 $\mu$m. Moreover, we observe the particle count increase towards lower size with addition of surfactant. Adsorption of nonionic surfactant molecules on the carbon black particle surface can lead to their better dispersion. The hydrophobic octylphenyl group would attach to the carbon black surface and the hydrophylic oxyethylene chain would extend outwards into the aqueous medium, thereby forming a steric layer on the particle surface and inhibiting particle aggregation.\cite{ma1992mixed,bossolelti1995adsorption,gupta2005adsorption}

\subsection{\label{level3.2}Effect of pH on surfactant CMC}
\begin{figure}
	\centering
	\includegraphics[width=0.8\linewidth]{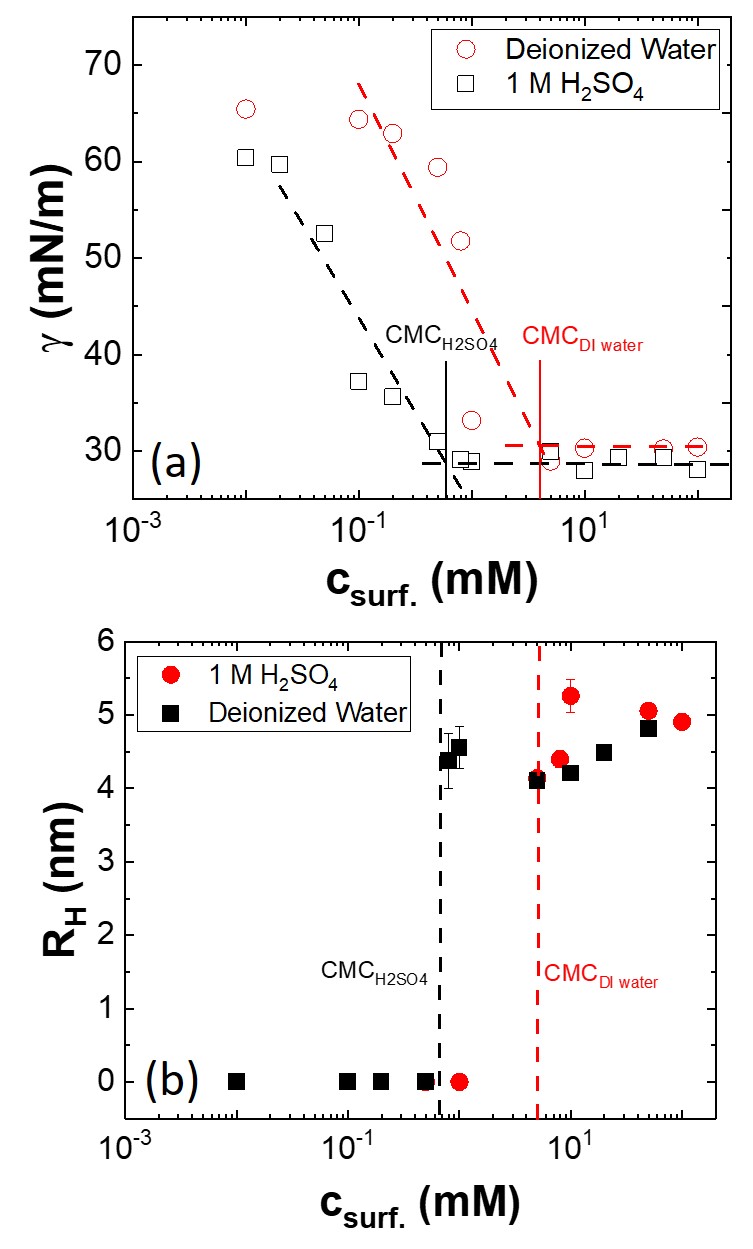}
	\caption[Fig4]{(a)Surface tension ($\gamma$) measured using pendant drop method shows the drop in CMC of the surfactant in 1 M H$_2$SO$_4$ compared to deionized water.(b) DLS measurement comparing the R$_H$ of micelles in deionized water and 1 M H$_2$SO$_4$.}
	\label{fig:fig4}
\end{figure}

Surface tension measurements reveal that the CMC of the surfactant dropped in 1 M H$_2$SO$_4$ by a decade compared to deionized water (Figure \ref{fig:fig4} (a)). In addition, we measured the hydrodynamic radius (R$_H$) of the micelles in both mediums using DLS. Our results suggest that both methods reveal similar values for CMC, however, the micelle size in 1 M H$_2$SO$_4$ appears slightly lower than in deionized water (Figure \ref{fig:fig4} (b)). Micelle size determined in deionized water here matches that reported in literature.\cite{charlton2000electrolyte} The nonionic surfactant monomer of Triton X-100 would form a `reverse micelle' in aqueous medium whereby the octylephenyl head group forms the hydrophobic core and the oxyethylene chain forms the hydrophilic surface.\cite{robson1977size} The hydrophilic chain of the nonionic surfactant molecule is more hydrated at high pH whereas the hydrocarbon core is less affected by changes in pH and would result in a larger R$_H$ for the micelle in deionized water.\cite{bloor1970effect} Moreover, Triton X-100 is expected to form spherical micelles close to CMC and oblate micelles above CMC that increase in size with surfactant monomer concentration.\cite{robson1977size,paradies1980shape,brown1989static}
\par For a suspension of carbon black particles dispersed in aqueous medium containing a nonionic surfactant, we expect the octylepheyl group to attach itself to the particle surface. Note that activated carbon black such as the one used here will have hydrophilic surface groups that may influence surfactant adsorption.\cite{boehm1994some} Scattering studies have shown that the surface coverage of the non activated carbon black particles by Triton X-100 is only 8\% and remains constant with increasing surfactant concentration.\cite{garamus1998small} It is shown that CMC drops with the addition of carbon black particles whereas the micelle size remains constant with and without the presence of particles. The reason for this is stated as the decreased solubility of surfactant molecules as a result of addition of hydrophobic carbon black particles. Furthermore, it is shown that Triton X-100 molecules formed a monolayer (thickness 32 A$^{\circ}$) on the carbon black particle surface and the layer thickness reduced (to 3 A$^{\circ}$) with the increase in surfactant monomer concentration. This is attributed to a change in orientation of surfactant molecule with respect to the particle surface from perpendicular to parallel.\cite{garamus1998small} A similar study is needed to establish the nature of Triton X-100 monomer adsorption on activated carbon black surface at low pH conditions. 

\subsection{\label{level3.3}Adsorption Isotherm}
\begin{figure}[h]
	\centering
	\includegraphics[width=0.8\linewidth]{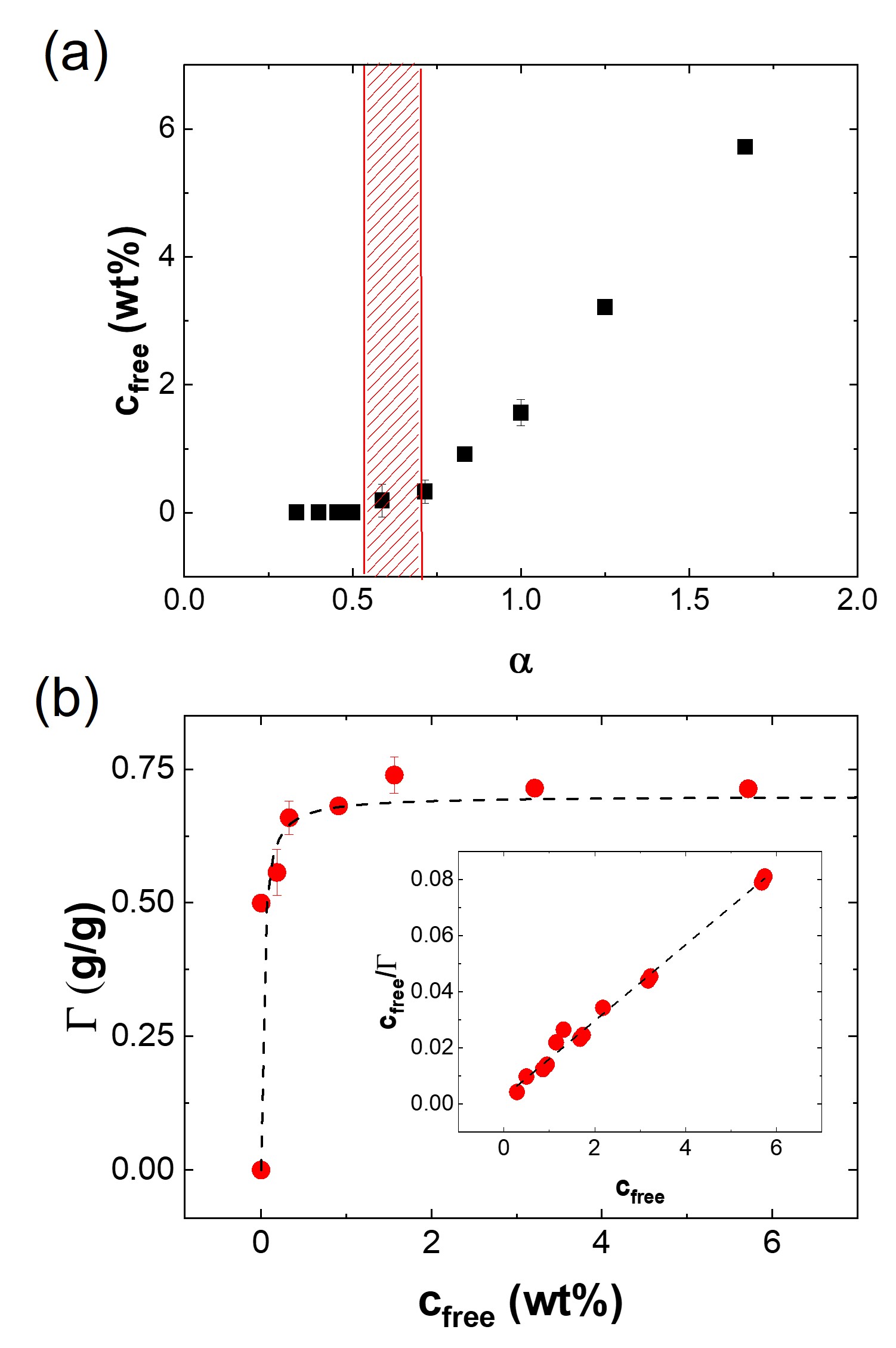}
	\caption [Fig5] {(a) Concentration of free surfactant remaining in the supernatant plotted against $\alpha$ (b) Langmuir adsorption isotherm (inset: linearized form) of the surfactant. Dashed line is the Langmuir adsorption fit with parameters obtained using Eq. \ref{eq:eq1} for the experimental results.}
	\label{fig:fig5}
\end{figure}

The free surfactant concentration c$_{free}$ in the supernatant is plotted for each sample with varying $\alpha$, with an initial c$_{surf.}$ = 0.06 to 0.1 g/ml at c$_{CB}$ = 6 g/ml ($\phi$ = 0.03) (Figure \ref{fig:fig5} (a)). At $\alpha$ < 0.55, there is little to no free surfactant in the supernatant, however, as $\alpha$ is increased above 0.55 we observe the presence of free surfactant. The shaded region in Figure \ref{fig:fig5} (a) ($\alpha$ = 0.55 - 0.7) is the regime where excess surfactant begins to appear in the supernatant. This is indicative of the saturation of the carbon black surface by surfactant molecules.
Assuming that the adsorbed surfactant forms a monolayer on the carbon black surface, we analyzed our results using a linearized form of Langmuir adsorption model given by Eq. \ref{eq:eq1}:
	\begin{equation}
		\dfrac{c_{free}}{\Gamma} = \dfrac{c_{free}}{\Gamma_{max}} + \dfrac{1}{\Gamma_{max}K}
		\label{eq:eq1}
	\end{equation}
where c$_{free}$ is the concentration of free surfactant in the supernatant, $\Gamma$ is the amount of surfactant adsorbed at the interface (and it's maximum, $\Gamma_{max}$), and \textit{K} is the equilibrium constant that describes the binding affinity of the adsorbed surfactant.

\par By plotting \textit{$\dfrac{c_{free}}{\Gamma}$} against $c_{free}$ as shown in Figure \ref{fig:fig5} (b) inset, we obtained values for $\Gamma_{max}$ and \textit{K} from the slope (\textit{$\dfrac{1}{\Gamma_{max}}$}) and the intercept (\textit{$\dfrac{1}{\Gamma_{max}K}$}) of the linear fit. The values of $\Gamma_{max}$ and \textit{K} are 0.7 g/g (1.1 $\times$ 10$^{-3}$ mol/g) and 37, respectively.  Consequently the area occupied per molecule was calculated using:\cite{gonzalez2000determination}
	\begin{equation}
	A = \dfrac{S_{N2}}{\Gamma_{max} N_A}
	\label{eq:eq2}
\end{equation}
where S$_{N2}$ is the BET surface area of carbon black (1600 m$^2$/g) and N$_A$ is Avogadro's number. Area occupied by surfactant was found to be 2.37 nm$^{2}$/molecule.	It needs to be noted here that the surface area available for surfactant molecules is much lower than N$_2$ due to former's larger molecular size. Our results are close to the values reported in literature for adsorption of Triton X 100 on carbon black.\cite{gonzalez2000determination,gupta2005adsorption} Adsorption study suggests that the surfactant molecules saturate the carbon black surface at $\alpha >$ 0.55 as indicated by the plateau in the adsorption isotherm (Figure \ref{fig:fig5} (b)).
	
\par Because of our interest in investigating sedimentation as a function of carbon black volume fraction, we fixed the surfactant concentration to 3 and 6 wt \% (or 0.05 and 0.1 M). This allows us to vary $\alpha$ below the saturation limit found using adsorption measurements.

\subsection{\label{level3.4}Gel collapse}
\begin{figure}[h]
	\centering
	\includegraphics[width=0.9\linewidth]{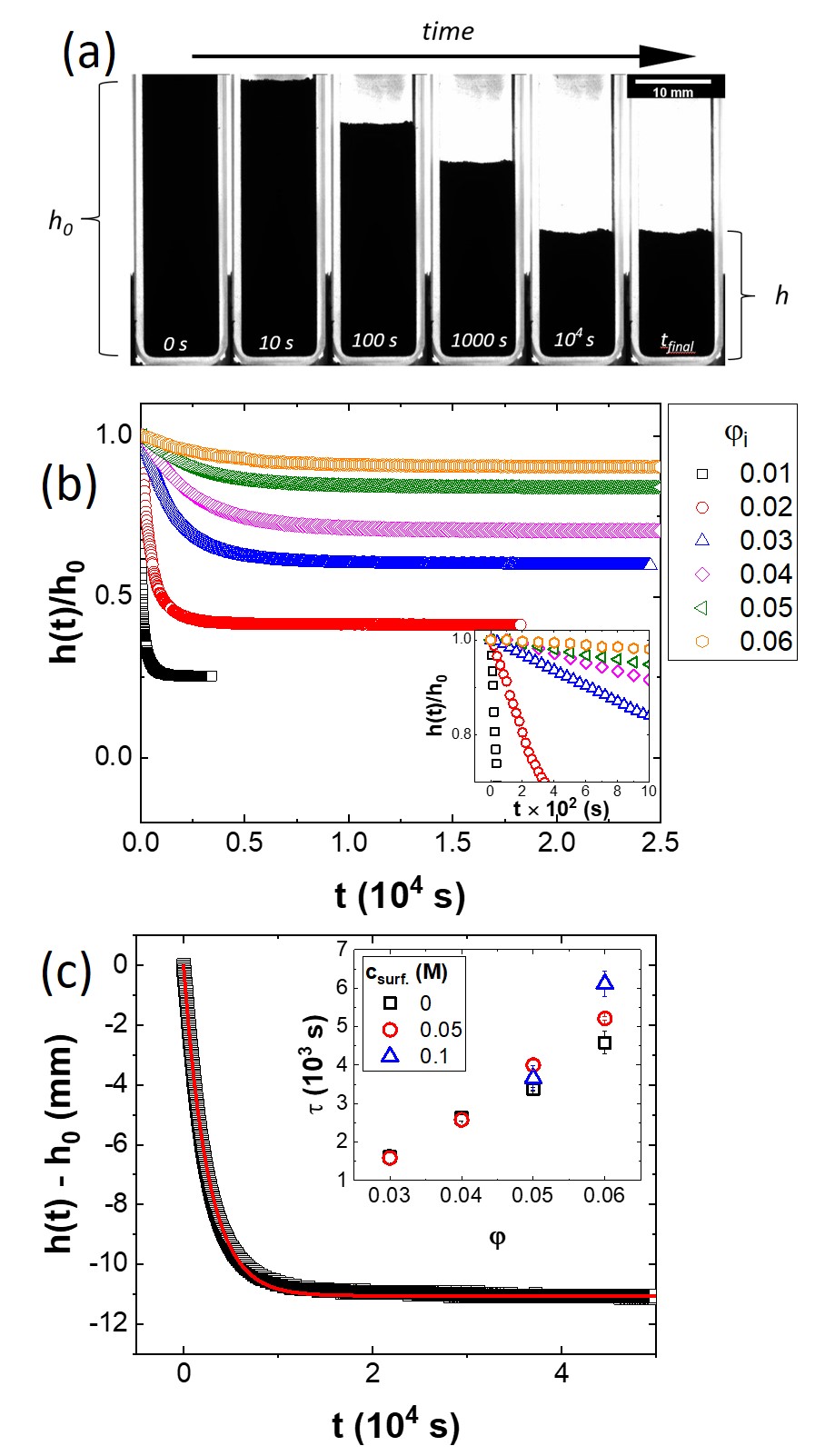}
	\caption[Fig6]{(a) Snapshots of particle sedimentation of a $\phi = 0.02$ carbon black suspension in a quartz cuvette, (b) temporal evolution of the normalized bed height (\textit{h/h$_0$}) for different volume fractions of carbon black dispersed in 1 M H$_2$SO$_4$, (c) red line: exponential fit from Eq. \ref{eq:eq1} for particle bed height of carbon black suspension ($\phi$ = 0.04), inset: values of $\tau$ obtained for different $\phi$.}
	\label{fig:fig6}
\end{figure}

Figure \ref{fig:fig6} (a) shows time-lapse images of sedimentation of a suspension of carbon black particles ($\phi$ = 0.04) dispersed in 1 M H$_2$SO$_4$. In suspensions without surfactant, as the particle bed sediments a clear supernatant devoid of any floating particles is observed. This allows us to clearly define the fluid-solid interface identified as the time dependent bed height, \textit{h (t)} during post-processing of the captured images. Moreover, it can be clearly seen here that the particles do not stick to the walls of the quartz cuvette. The particle bed height (normalized by initial height, \textit{h$_0$} $\approx$ 36.5 $\pm$1 mm) is plotted as a function of time for different carbon black volume fractions. 
\par Carbon black particles when dispersed in aqueous mediums have a negative surface charge.\cite{xu2007particle} However, when dispersed in 1 M H$_2$SO$_4$, the magnitude of their surface charge reduces as a result of excess hydronium counter-ions in the low pH suspending medium. Despite the presence of hydrophilic surface groups on activated carbon black particles, they will form aggregates at low pH in polar mediums in absence of strong electrostatic double layer repulsion and driven by van der Waals attraction. Attractive colloidal particles (here carbon black particles) at sufficiently low $\phi$ known as critical point (or $\phi*$) can form a space spanning percolated network leading to an out-of-equilibrium metastable state known as a gel.\cite{krall1998internal, lu2008gelation,zaccarelli2007colloidal} A gel has a high viscosity and a finite yield stress. As the carbon black suspension is allowed to stand, the density mismatch between the particles and the suspending medium will lead to sedimentation, at the same time interparticle attraction would lead to gel formation. However, the yield stress behavior of the resultant gel is expected to either deter or slow down this process. Understanding the mechanism of sedimentation is crucial in improving the gravitational stability of carbon black suspensions used in electrochemical applications.
\par Sedimentation in colloidal gels is governed by a balance between gravitational stress, network elastic stress and viscous drag on the fluid arising from network porosity. Depending on the range and strength of interparticle attraction \textit{U}, the particle volume fraction $\phi$ and the height of the container, sedimentation in gels can exhibit different mechanisms.\cite{starrs2002collapse,manley2005gravitational,condre2007role,harich2016gravitational} For gels comprised of particles with short range attraction, at high $\phi$ the particle bed follows a steady or ``creeping'' flow in agreement with Darcy flow.\cite{manley2005gravitational} However, at lower $\phi$ and/or low \textit{U} the sedimentation typically proceeds in three stages. In the first stage, the gel exhibits a very slow creeping sedimentation, followed by the second stage characterised by a rapid loss of bed height known as `gel collapse' and lastly, the third stage whereby a very slow compaction of the sediment takes place.\cite{allain1995aggregation,poon1999delayed,starrs2002collapse,manley2005gravitational,condre2007role,buscall2009towards} It is known that the first stage is influenced either by sedimentation of individual aggregates or by the back flow of the fluid through the particle bed due to a no flux boundary condition at the bottom wall of the container. Hence, here the fluid flow is hindered by the network porosity and is very slow.\cite{russel1991colloidal,buscall1987consolidation} In some cases it has been shown that gels exhibit ``delayed collapse'' after a delay time $\tau_d$ which is related to the coarsening of the gel network due to localized particle rearrangements.\cite{poon1999delayed,gopalakrishnan2006linking,lietor2010role,bartlett2012sudden} The second stage of gel collapse is influenced by internal structural rearrangements driven by interparticle aggregation compounded by fluid flow under gravity. As a result, different mechanisms have been shown to induce structural failure such as formation of fluid channels or streamers within the fractured gel\cite{allain1995aggregation,poon1999delayed,derec2003rapid} and ageing induced particle clustering resulting in localized ``microcollapses''.\cite{bartlett2012sudden} It needs to be noted that in gels with strong interparticle attraction such as the one in this study, localized structural rearrangement driven by particle Brownian motion are extremely slow compared to the time scale of sedimentation.\cite{buscall2009towards} Third and the final stage is the compaction of the sediment where the balance between network elastic forces and gravitational stress as well as bed porosity at higher particle volume fraction determines the equilibrium conditions.\cite{manley2005gravitational,lee2006formation}

\par In the carbon black suspensions studied here the initial region of collapse of the particle bed (excluding the compaction stage) could be fit well using a exponential decay function (Eq. \ref{eq:eq2}) and poroelastic model as proposed by Manley et al.\cite{manley2005gravitational} for gel collapse given by:

\begin{equation}
	h_0 - h (t) = \Delta h(1- e^{-t/\tau})
	\label{eq:eq3}
\end{equation}
where \textit{h$_0$} is the intial particle bed height, the total change in bed height is given by $\Delta$h = $\dfrac{\Delta \rho g \phi h_0^{2}}{2E}$ and time scale associated with gel collapse in a solvent with viscosity $\eta$ is given by $\tau$ = $\dfrac{4\eta (1-\phi) h_0^{2}}{\pi^2 k_0 E}$. Here we can obtain \textit{E}, the elastic modulus and \textit{k$_0$}, the permeability of the gel. This model was proposed by Manley et al. for the gels of silica particles in the regimes where they did not observe fracture and leads us to believe that a similar behavior is exhibited by our carbon black gels. In our samples $\tau$, E and k$_0$ show a linear relation with suspension $\phi$ (Figure S5). 

\begin{figure}[h]
	\centering
	\includegraphics[width=0.8\linewidth]{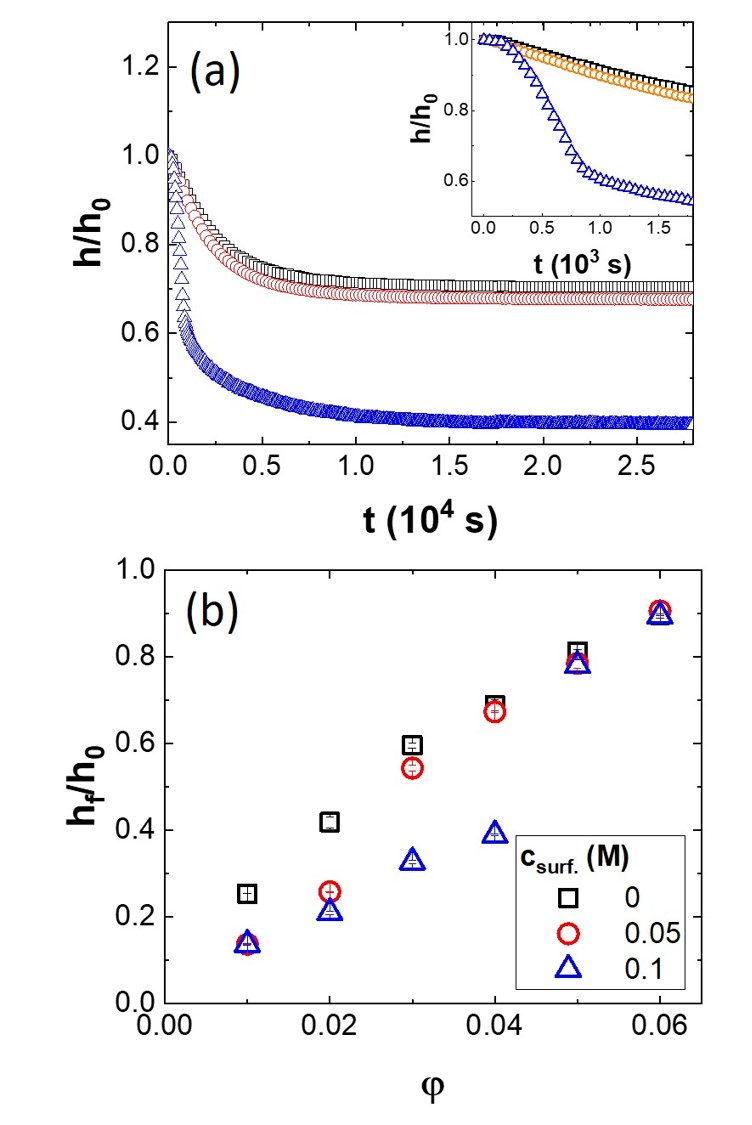}
	\caption[Fig7]{(a) Normalized bed height (\textit{h/h$_0$}) plotted against time for  $\phi = 0.04$ carbon black dispersed in 1 M H$_2$SO$_4$ with different surfactant concentrations, (inset: blow up of short time collapse) (b) normalized final bed height (h$_f$/h$_0$) for different particle volume fractions at different surfactant concentrations in 1 M H$_2$SO$_4$.}
	\label{fig:fig7}
\end{figure}

\textbf{Catastrophic collapse:} Addition of surfactant affects the sedimentation dynamics significantly. Firstly, at low particle volume fractions ($\phi \le$ 0.02) there exists a diffuse layer of particles above the sedimenting bed that obscures the fluid-solid interface necessary for visual investigation of sedimentation dynamics. Secondly, there is a significant increase in the sedimentation rate at short times in some of the samples leading to a `catastrophic collapse' (inset: Figure \ref{fig:fig7} (a)) and is similar to that reported for low volume fraction gels in tall sample cells.\cite{poon1999delayed,starrs2002collapse,senis1997scaling,manley2005gravitational} Moreover, the final bed height is lower in suspensions exhibiting particle diffuse layer as well as catastrophic collapse resulting in a higher packing factor (Figure S4, SI). It needs to be noted that none of our suspensions exhibit delayed collapse as has been observed in some other type of gels.\cite{poon1999delayed,starrs2002collapse,buscall1987consolidation,bartlett2012sudden} It is possible a regime of delayed collapse might exist at $\phi >$0.06 and is not explored in this study.
\par Our adsorption measurements revealed that having $\alpha >$ 0.55 ensures surfactant molecules would saturate the carbon black surface which will induce interparticle steric repulsion.  As a result this would suppress particle aggregation and lead to weakening or elimination of gel formation. Interestingly, our results show that for a \textit{$\alpha <$} 0.7, the suspensions do not exhibit a catastrophic collapse and do not have a higher packing fraction for the sediment. Whereas for \textit{$\alpha >$} 0.7, there exists a diffuse layer of particles during sedimentation as well as a higher packing fraction for the sediment. Under the same condition at 0.02 $<\phi<$0.05 we observe catastrophic collapse. It is possible that similar behavior is exhibited even at $\phi \le$ 0.02, however, a large diffuse layer of particles obscures the fluid-solid interface. Both adsorption and sedimentation studies are in agreement that at $\alpha \ge$ 0.55, the carbon black surface is sterically stabilized by surfactant molecules leading to a weaker gel. Saturation of most of the carbon black surface by surfactant molecules will leave little free surface to form interparticle bonds. This will lead to a reduced number of contact points between particle clusters and the resulting gel network will not be able to withstand gravitational stress. $\phi_{max} \approx$ 0.20 determined using centrifugation reveals that carbon black particles used here cannot be packed to very large $\phi$ (e.g. $\phi_{max} \approx$ 0.64 observed for hard spheres\cite{mewis2012colloidal}) due to the inherent anisotropy of the primary aggregates. However, a higher $\phi$ of the sediment for samples with \textit{$\alpha >$} 0.7 reveals that it is still below $\phi_{max}$ and may possess some elasticity.
\par Sedimentation dynamics revealed interesting similarities between our samples and that of weak colloidal gels. Moreover, we observed the role of surfactant on interparticle interaction affecting sedimentation dynamics and packing fraction of the sediment at equilibrium. However, in addition to gravitational stability, the viscoelastic response of gels at rest and during flow measured using linear and nonlinear rheology forms the basis of their final application. In our case, the rheological behavior of our samples is crucial in determining their use for slurry electrodes used in all-iron flow batteries.

\subsection{\label{level3.5}Gel viscoelasticity}
Rheological measurements were carried out to determine whether and at what conditions carbon black formulations formed a gel. We applied the highest shear rate achievable in the rheometer (1000 s$^{-1}$) for sample rejuvenation. High shear rejuvenation would break interparticle bonds as a result of viscous drag force being greater than interparticle attraction force and ensures particle dispersion.\cite{mewis2012colloidal,masschaele2011flow,varga2018large} This also subjects all samples to the same shear history and eliminates any preshear effects commonly encountered in such systems.\cite{mewis1979thixotropy,ovarlez2013rheopexy,osuji2008shear} 
\par
\begin{figure}
		\centering
	\includegraphics[width=0.9\linewidth]{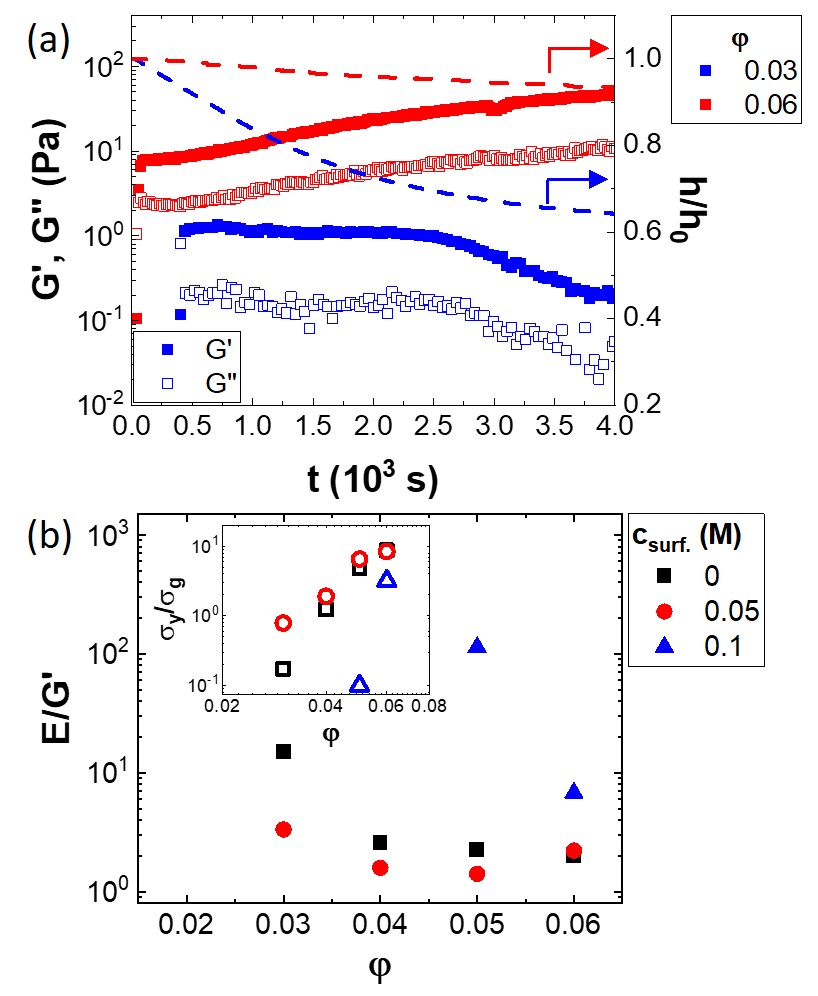}
	\caption[Fig8]{(a) Time evolution of viscoelasticity (ageing) after rejuvenation of carbon black suspensions with c$_{surf.}$ = 0 compared with temporal evolution of gel height in the cuvette (dashed curves). Drop in G' and G'' for  $\phi$ = 0.03 indicate a weakening of the particle network and probable sedimentation within the rheometer geometry. (b) Elastic modulus (\textit{E}) obtained from Eq. \ref{eq:eq1} normalized by storage modulus (G') obtained after ageing (at $\gamma$ = 0.001, $\omega$ = 1 rad s$^{-1}$, t$_w$ = 4000 s), inset: yield stress ($\sigma_y$) determined from DSS measurement normalized by gravitational stress ($\sigma_g$) for carbon black suspension at different c$_{surf.}$. }
	\label{fig:fig8}
\end{figure}

 \textbf{Linear rheological} measurements performed after sample rejuvenation show that for samples that do not exhibit catastrophic collapse, a time-evolution of viscoelasticity also known as ageing is observed. Figure \ref{fig:fig8} (a) shows typical ageing behavior exhibited by a carbon black suspension after flow cessation. The sample exhibits solid-like response where the storage modulus (\textit{G'}) is greater than the loss modulus (\textit{G''}). Moreover, the solid-like behavior is observed at very short waiting time t$_w$ ($<$ 100 s) indicating a rapid percolation of the particle network after rejuvenation. This is known as the sol-gel transition.\cite{koumakis2015tuning,das2021shear} At low particle volume fractions ($\phi \le$0.03), we observe a slight drop in solid-like behavior with time indicative of structural collapse under gravity. This is probably due to lower number of contact points within the gel network that results in faster inter-cluster bond breaking in the presence of gravity. There can also be additional contributions from network coarsening.\cite{teece2011ageing,bartlett2012sudden,harich2016gravitational} At higher particle volume fractions, effect of ageing is more prominent with \textit{G'} and \textit{G''} values increasing by an order of magnitude over time as is observed in gels formed by interparticle van der Waals attraction.\cite{bonacci2020contact,n2020yielding} However, during sedimentation study we observe gel collapse in all the samples (see Figure \ref{fig:fig6}) since the time scale associated with ageing is slower (as a result of strong interparticle attraction or \textit{U} $\gg$ \textit{k$_B$T}) than the time scale associated with sedimentation. 
\par Next we compared the values of elastic modulus (\textit{E}) obtained from poroelastic model with shear storage modulus (\textit{G'}). We found \textit{E} $\approx$ 2\textit{G'} except at $\phi$ = 0.03 for c$_{surf.}$ = 0 and 0.05 M. Moreover, values of \textit{G'} were slightly higher for suspensions containing c$_{surf.}$ = 0.05 M. One may expect that addition of surfactant would always lead to a reduction in \textit{G'} compared to no surfactant condition due steric repulsion between carbon black particles induced by adsorbed surfactant molecules.\cite{khalkhal2018evaluating} However, this is not the case here. Further increasing c$_{surf.}$ to 0.1 M results in liquid-like behavior at $\phi \le$ 0.05 and large drop in \textit{G'} values. We determined the yield stress, $\sigma_y$ of these gels using dynamic strain sweep (DSS) measurements (Figure S6, SI). $\sigma_y$ is determined at strain (known as yield strain) at which maximum dissipation of elastic stress (\textit{G'$\gamma$}) is observed. Beyond $\sigma_y$ the particle network undergoes irreversible deformation before the interparticle bonds rupture and the gel yields.\cite{walls2003yield} In the presence of gravity, the yield stress of a gel should be at least equal to the gravitational stress exerted by a single particle (or carbon black aggregate here) of radius \textit{a} within the gel network given by,\cite{harich2016gravitational}
\begin{equation}
	\sigma_g = \frac{4}{3}\Delta \rho g a \approx 11.3 \quad mPa
	\label{eq:eq4}
\end{equation}
In suspensions with c$_{surf.}$ = 0 and 0.05 M, we found that $\sigma_y \ge \sigma_g$ (except at $\phi$ = 0.03) and they exhibit gel collapse. However, at c$_{surf.}$ = 0.1 M we see considerable weakening of the gel and were unable to perform rheological measurements for $\phi \le$ 0.05 as the stresses in these suspensions could not be detected by the instrument. Despite having very low values of \textit{G'} and $\sigma_y$, $\phi \ge$ 0.05 show values of $\tau$ similar to the suspensions with c$_{surf.}$ = 0 and 0.05 M (Figure \ref{fig:fig6} (c) inset) indicating a similar sedimentation behavior. Moreover, the influence of increasing gel elasticity with $\phi$ is also evident in the final equilibrium height of the sediment \textit{h$_f$} as shown in Figure \ref{fig:fig7} (b) where \textit{h$_f$}/\textit{h$_0$} is approaching 1 for $\phi \ge$ 0.06. 

\par
\begin{figure}
	\centering
	\includegraphics[width=0.75\linewidth]{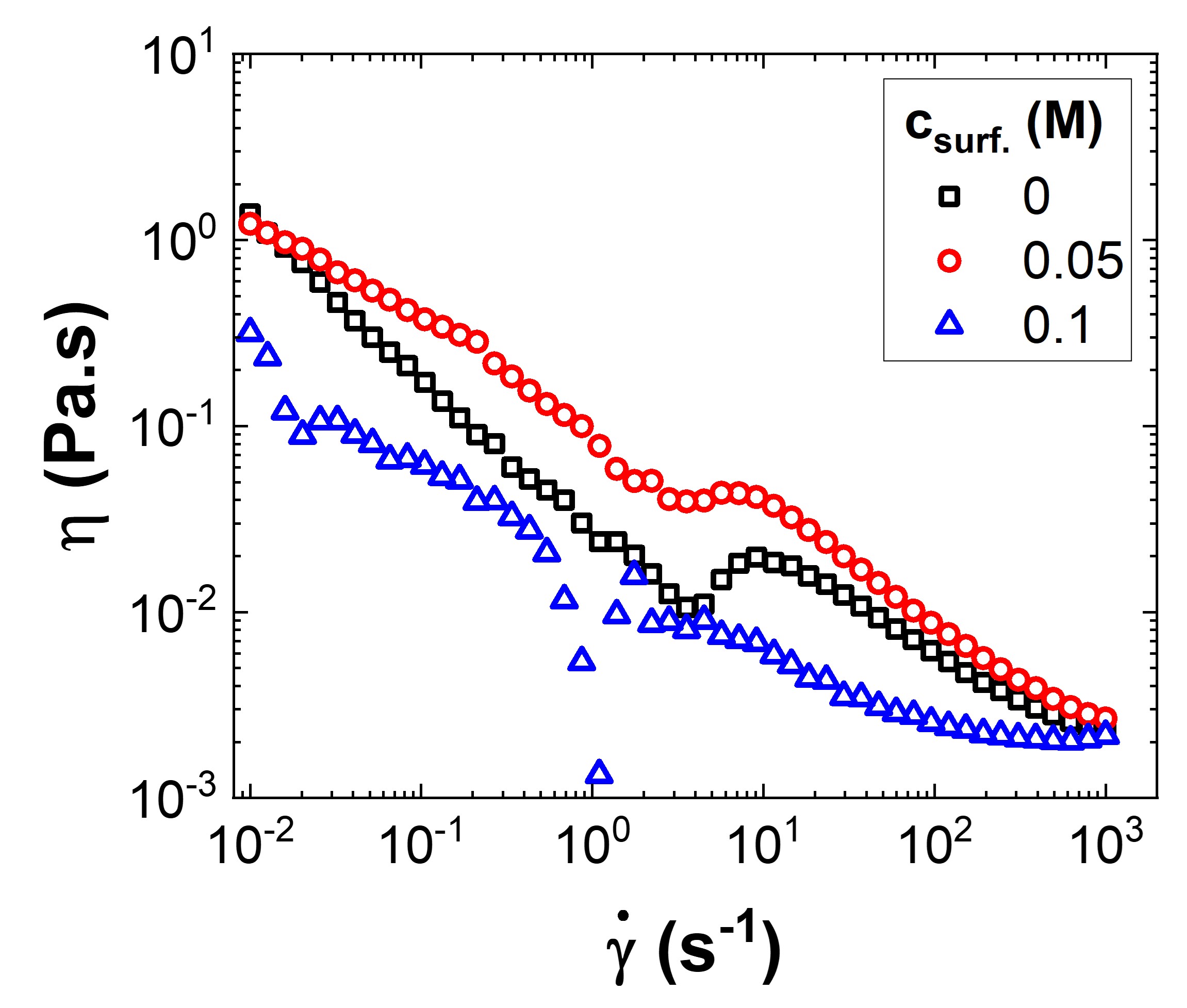}
	\caption[Fig9]{Measured viscosity ($\eta$) plotted against shear rate ($\dot{\gamma}$) for carbon black suspension ($\phi$ = 0.04) at different c$_{surf.}$.}
	\label{fig:fig9}
\end{figure}

 \textbf{Steady shear} measurements reveal that the carbon black suspensions exhibit shear thinning behavior. In Figure \ref{fig:fig9} we compare the flow curve of carbon black suspension ($\phi$ = 0.04) at different c$_{surf.}$. We see that at c$_{surf.}$ = 0.05 M there is a slight increase in suspension viscosity whereas at c$_{surf.}$ = 0.1 M, the viscosity drops drastically. Another important observation is a shear thickening regime where viscosity $\eta$ increases at intermediate shear rates $\dot{\gamma}$. This behavior is ubiquitous in carbon black and many other attractive particle suspensions.\cite{osuji2008shear,negi2009new,das2021shear} This is a result of shear induced structuring (at low $\phi$) commonly associated with such systems. At low shear rates we observe shear thinning as the percolated network of particles break down into large aggregates. Due to confinement effects within the small gap of the rheometer geometry and competition between interparticle attraction force and viscous drag force, the aggregates form mesoscopic particle rich flocs.\cite{relshear} These flocs break down into smaller aggregates at intermediate shear rates as hydrodynamic forces dominate. The resulting increase in hydrodynamic volume of the aggregates leads to an increase in suspension viscosity as indicated by the kink shown in Figure \ref{fig:fig9}. At higher shear rates, these aggregates break down into individual particles resulting in further shear thinning. From Figure \ref{fig:fig9} it is clear that the aggregation process during flow is disrupted by the addition of surfactant leading to a lower viscosity and shear thinning behavior of the suspension.  

\par Based on both linear and nonlinear rheological measurements it can be said at c$_{surf.}$ = 0.05 M, there is an increase in particle dispersion which increases the elasticity as well as viscosity of the gel compared to the no surfactant condition. Hence, at low c$_{surf.}$ (or \textit{$\alpha <$} 0.7) the surfactant driven increase in particle dispersion leads to a stronger gel. However, further increasing the c$_{surf.}$ (or \textit{$\alpha >$} 0.7) leads to a complete suppression of interparticle aggregation resulting in lower viscoelastic behavior as well as leads to a catastrophic gel collapse.

\section{\label{level4}Conclusion}
We studied the influence of nonionic surfactant Triton X-100 on gravitational stability and rheology of suspensions of carbon black particles dispersed in 1 M H$_2$SO$_4$ used in RFBs. Our studies revealed that the suspensions form aggregates due to interparticle van der Waals attraction and exhibit gel collapse behavior when allowed to stand. Using time-lapse imaging we determined the sedimentation kinetics revealing different collapse mechanisms as a result of surfactant addition. Addition of surfactant above CMC leads to better particle dispersion. and slight increase in gel elasticity and viscosity. However, at very high c$_{surf.}$, interparticle attraction is completely suppressed and the gel undergoes catastrophic collapse with denser packing for the sedimented bed. We determined that the value of $\alpha$ (= \textit{c$_{surf.}$/c$_{CB}$)} below 0.7 improves particle dispersion, suspension elasticity and prevents catastrophic gel collapse. Moreover, it can be concluded that the suspension sedimentation can be minimized at  $\alpha \ll$ 0.7 for the height used in this study. Our study is relevant in understanding the gravitational stability of slurry electrodes used in RFBs and the role of nonionic surfactant in improving particle dispersion and flowability, both of which have implications for battery performance. Additionally, results from our study can be useful for investigating the role of surfactants on sedimentation in other attractive particle suspensions. We plan to extend this study to understand the influence of container dimensions as well as particle volume fraction on gravitational stability of carbon black suspensions.
		
\begin{acknowledgments}
The authors acknowledge the financial support from Department of Energy, Office of Electricity, Pacific Northwest National Laboratory, USA (Contract No. 540358). The authors thank Joao Maia (Center for Advanced Polymer Processing, Case Western Reserve University) for providing access to the rheometer and Jesse Wainright and Bob Savinell (Case Western Reserve University) for technical discussions.
\end{acknowledgments}

\section{\label{level5}Supplementary Information}
\renewcommand{\thefigure}{S\arabic{figure}}

\setcounter{figure}{0}

\subsection{Effect of mixing protocol on particle size distribution}
Carbon black suspensions are extremely sensitive to preparation methods.\cite{akuzum2017effects} We explored different protocols for dispersing carbon black particles in 1 M H$_2$SO$_4$. Primarily we investigated the role of mechanical stirring, sonication, homogenization and a combination of all methods on the particle size distribution. The sample was stirred for more than 12 h. Following this, the sample was subjected to different dispersion protocols. For each protocol a part of the sample obtained after stirring was used. 
\begin{figure}[h]
	\centering
	\includegraphics[width=1\linewidth]{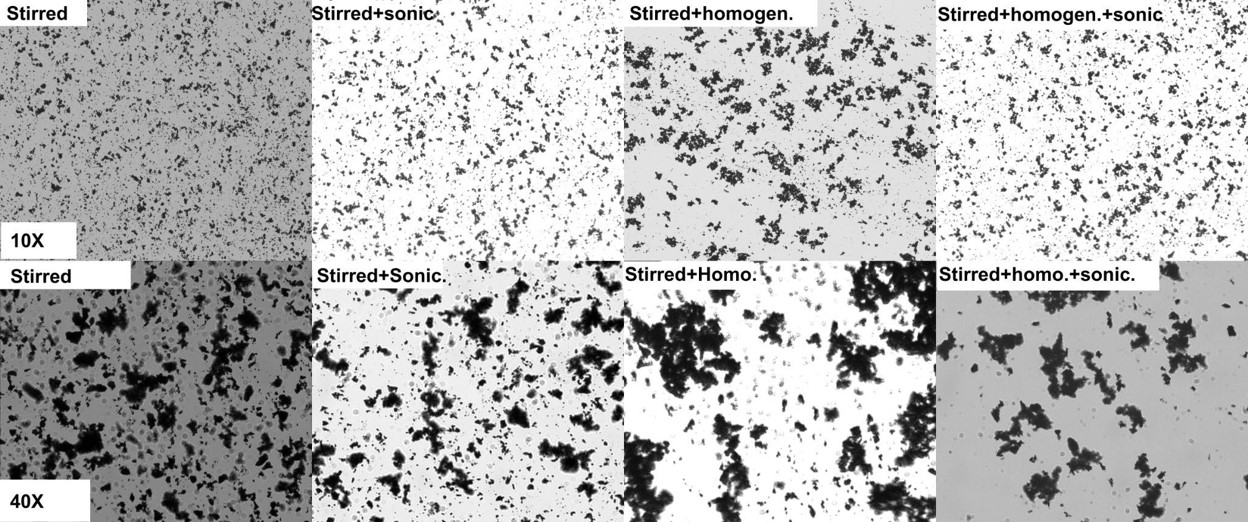}
	\caption[Figure S1]{Microscopy images of carbon black suspension in 1 M H$_2$SO$_4$ ($\phi$ = 0.01) after different mixing protocols.}
	\label{fig:figs1}
\end{figure} 
\par
\begin{figure}[h]
	\centering
	\includegraphics[width=0.7\linewidth]{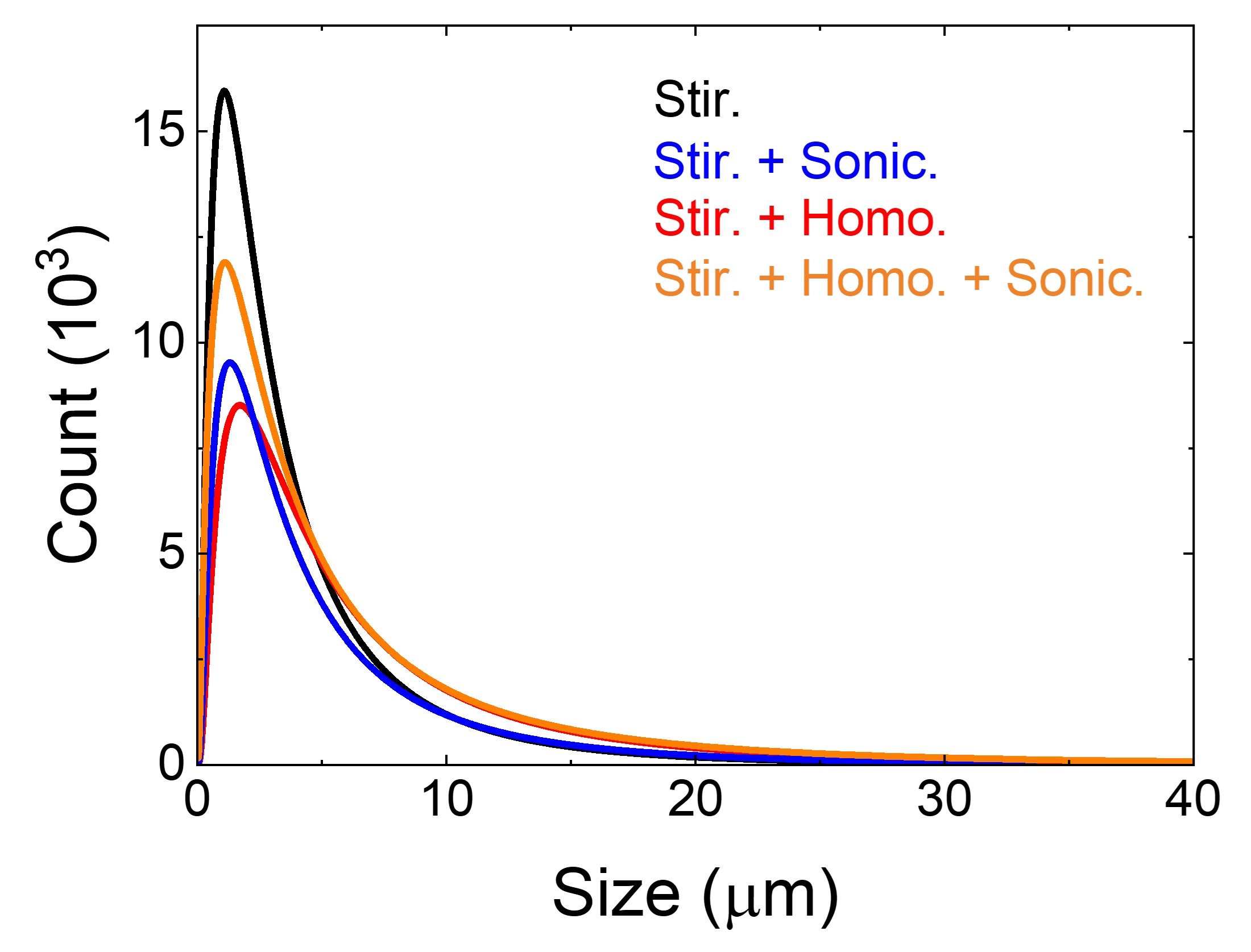}
	\caption[FigS2]{Log normal distribution of particle size obtained after analyzing microscopy images of samples prepared using different dispersion protocols.}
	\label{fig:figs2}
\end{figure}

 Figure \ref{fig:figs1} shows microscopic images of carbon black suspensions at $\phi$ = 0.01 captured at 10$times$ and 40$times$ magnification. It is clear that after different protocols, we still observe large particle aggregates. However, there were some unique observations as shown in Figure \ref{fig:figs2}. Stirring process introduced the widest particle distribution indicating the strongest effect on particle dispersion. It needs to be noted that when dispersed in very low pH mediums, carbon black particles would tend to aggregate. However, activated carbon black has more hydrophylic surface groups and hence maybe easier to disperse in aqueous mediums such as used here just by simple stirring. Stirring followed by sonication lead to a shift in distribution peak to larger size with the appearance of larger particle aggregates. A likely explanation for this behavior could be the strong van der Waals attraction forces between carbon black particles dominating the cavitation induced break up of interparticle bonds. Next we used a homogenizer on a stirred sample for 5 min which introduced shear rates in the range of several 1000 s$^{-1}$. However, we observed largest particle aggregates using this protocol contrary to the expectation that high shear rates can induce break up of particle aggregates and lead to better dispersion. It is possible that there were several dead-zones within the sample container during the homogenization process which could have lead to increased aggregation. A combination of all the methods lead to smaller aggregates compared to homogenization, however, did not lead to better dispersion compared to mechanical stirring. Hence, for all our sample preparation we used mechanical stirring process.

%
\subsection{Image analysis - Sedimentation kinetics}
We measured the sedimentation rate by tracking the bed height of the carbon black particle suspension as it sediments with time. A custom optical set up comprising of a CMOS camera (CS235MU, Thorlabs) equipped with a zoom lens (6.5$\times$, Thorlabs), quartz cuvette (CV10Q35, Thorlabs) and a white light LED illumination source was used. Images were recorded at 0.1, 0.25 and 1 fps depending on the total duration of the experiment. Image analysis to identify fluid-solid interface was carried out in the following steps:
\begin{itemize}
	\item 8-bit images were captured using time-lapse imaging
	\item Image grayscale (GS) values vs. cuvette height (H) were plotted along 20 locations along the cuvette width. As the fluid-solid interface was almost parallel to the cuvette bottom wall, 20 locations were sufficient for identifying the interface.
	\item Number average filter algorithm in MATLAB was used to eliminate noise and to smoothen the curves
	\item Calculated the change in GS values along H (dGS/dH) and plotted them against H
	\item Located the maximum of dGS/dH curve and defined it as height of the interface (\textit{h}). This was done to account for the formation of a diffuse layer of particles in the fluid-solid interface during sedimentation.
	\item Repeat the process for all the images captured over time during sedimentation.
\end{itemize}

\begin{figure}[h]
	\centering
	\includegraphics[width=1\linewidth]{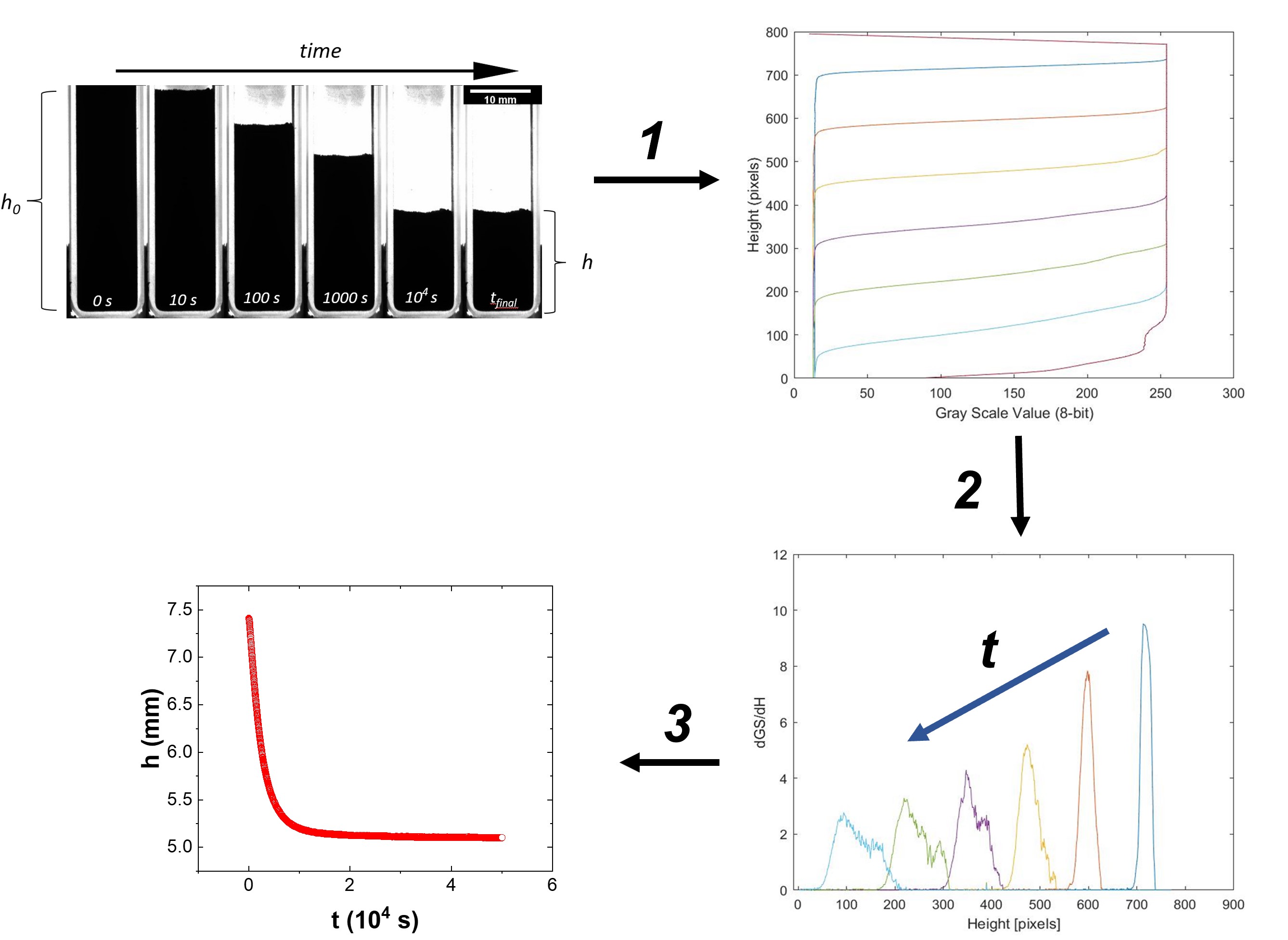}
	\caption[FigS3]{Image processing steps followed to determine fluid-solid interface for a sedimenting particle bed. Different colors in MATLAB plots (second and third) represent grayscale values captured at different times as the particle bed sediments.}
	\label{fig:figs3}
\end{figure}

\subsection{Maximum packing}
\par
\begin{figure}[h]
	\centering
	\includegraphics[width=0.75\linewidth]{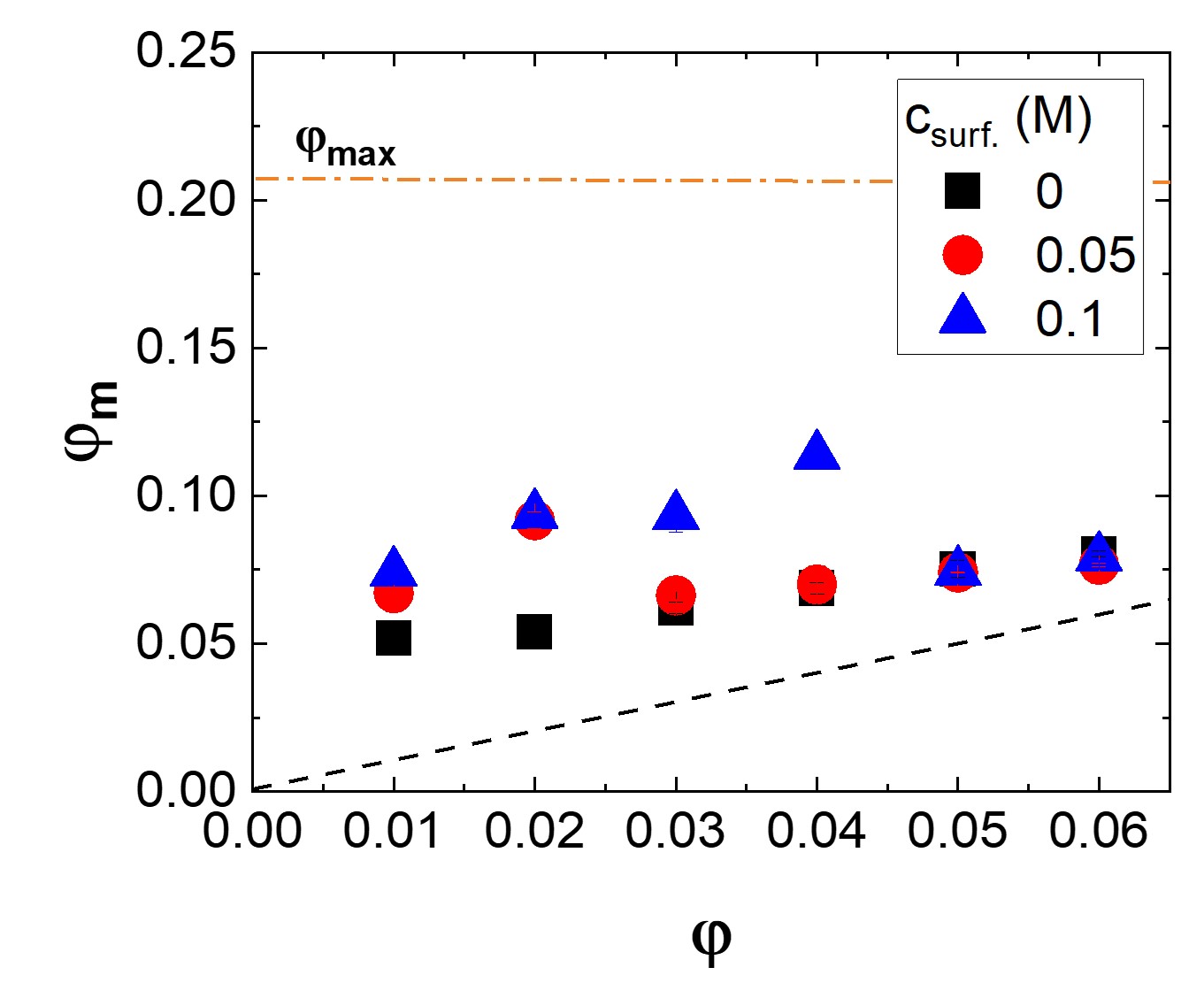}
	\caption[FigS4]{Maximum packing fraction ($\phi_{m}$) determined for sedimented bed of carbon black particles at different surfactant concentrations. Dashed line indicates $\phi = \phi_{m}$ and dash-dot line is the random close packing limit ($\phi_{max}$) determined from centrifugation of $\phi$ = 0.05 suspension at 7000$\times$g. Suspensions exhibiting catastrophic collapse and diffused layer of particles have a higher packing fraction after sedimentation.}
	\label{fig:figs4}
\end{figure}

 We determined the maximum packing ($\phi_{m}$) of the carbon black suspensions after sedimentation experiment (Figure \ref{fig:figs4}). The clear supernatant was removed using a micropipette (100$\mu$l at a time) without disturbing the particle bed. The volume fraction of the carbon black in the sediment was calculated from mass of the sample before and after drying in a hot air oven at 100 $^{\circ}$C for 48 h and after accounting for the mass of non-evaporated H$_2$SO$_4$. Our values followed the conservation of volume of carbon black particles, \textit{$\phi$ = $\phi_0$h$_0$/h}.

\subsection{Elastic modulus and permeability}
\begin{figure}[h]
	\centering
	\includegraphics[width=0.8\linewidth]{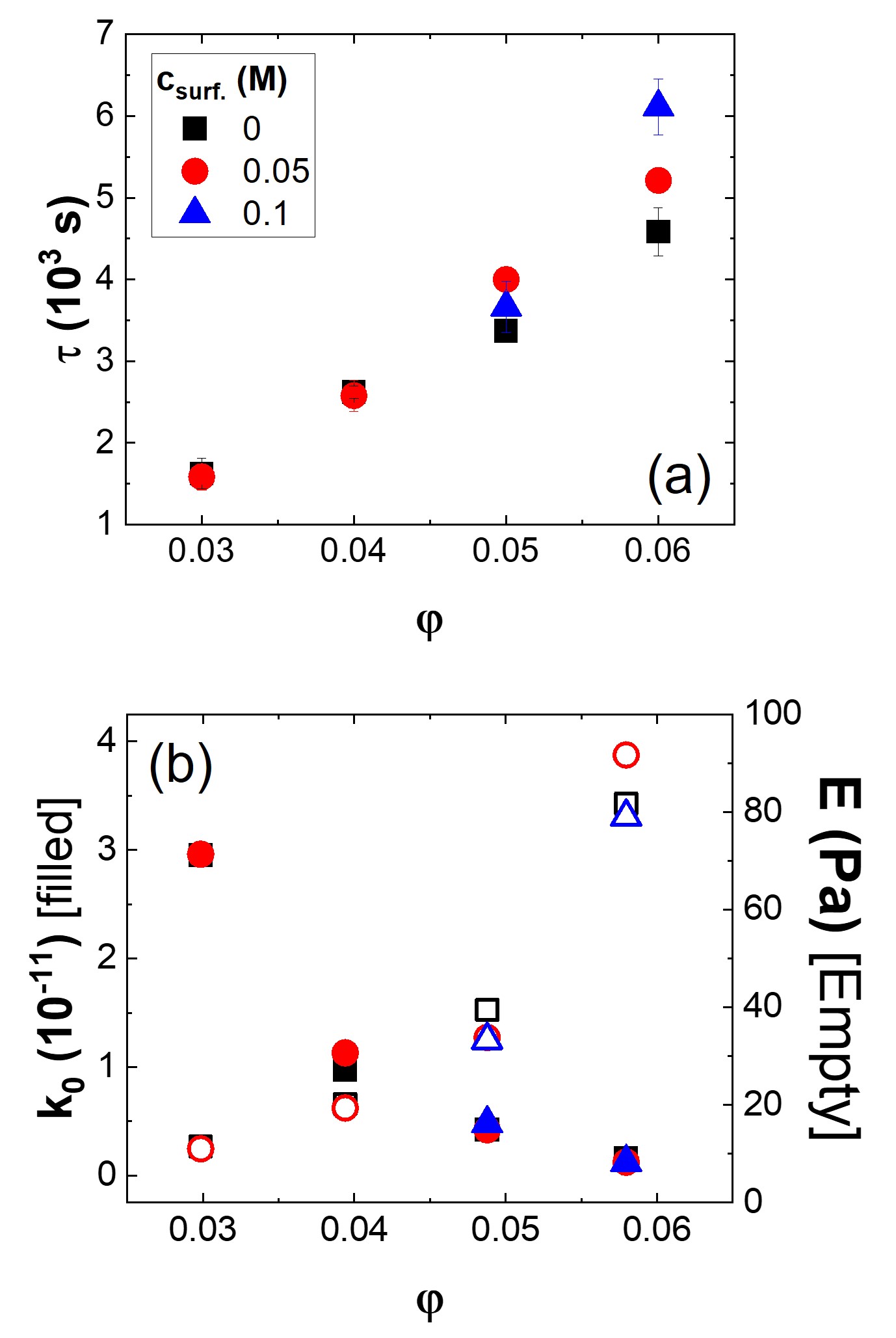}
	\caption[FigS5]{(a) Time scale of gel collapse $\tau$ obtained from fitting Eq. \ref{eq:eq3} and (b) elastic modulus \textit{E} and gel permeability k$_0$ obtained from poroelastic model plotted as function of carbon black volume fraction $\phi$ at different surfactant concentrations.}
	\label{fig:figs5}
\end{figure}

Sedimentation kinetics in our samples where catastrophic collapse wasn't observed, followed a exponential decay proposed by Manley et al.\cite{manley2005gravitational} based on poroelastic model. We were able to extract the values of elastic modulus \textit{E} and gel permeability \textit{k$_0$} from the model and is shown in Figure \ref{fig:figs5}.

\subsection{Ageing and yielding}
\begin{figure}[h]
	\centering
	\includegraphics[width=0.8\linewidth]{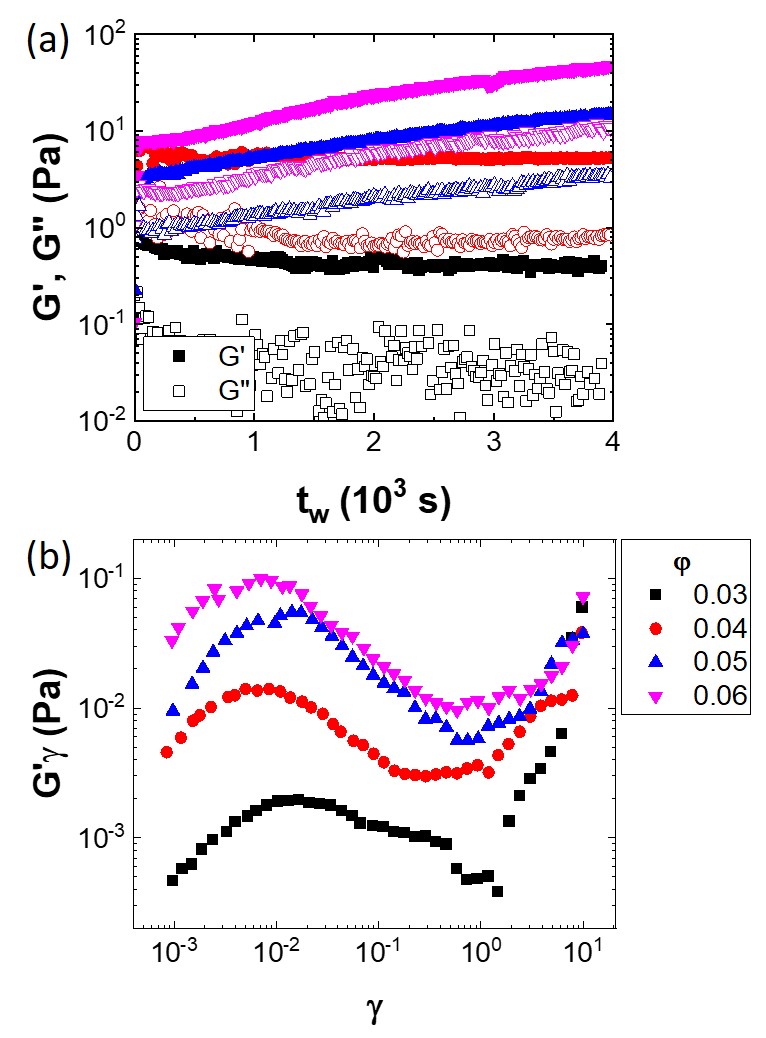}
	\caption[FigS6]{(a) Ageing of carbon black gel after high shear rejuvenation, (b) Elastic stress (G'$\gamma$) plotted against $\gamma$ for carbon black gels at different particle volume fractions with c$_{surf.}$ = 0. The first maximum of G'$\gamma$ represents the yield stress under oscillatory shear.}
	\label{fig:figs6}
\end{figure}

In addition to studying ageing, we measured the yield stress ($\sigma_y$) of the carbon black suspension undergoing oscillatory shear using dynamic strain sweep (DSS) measurement. Yield stress in gels is determined at the strain amplitude at which maximum dissipation of elastic stress (\textit{G'$\gamma$}) is observed when the particle network undergoes irreversible deformation as strain is increased before the interparticle bonds rupture and the gel yields.\cite{walls2003yield} Figure \ref{fig:figs6} shows ageing and yielding behavior of carbon black suspensions that exhibit gel collapse in the absence of surfactant.

\section{References}		
		\nocite{*}
		\bibliography{references}

\begin{thebibliography}{10}

\bibitem{dunn2011electrical}
Bruce Dunn, Haresh Kamath, and Jean-Marie Tarascon.
\newblock Electrical energy storage for the grid: a battery of choices.
\newblock {\em Science}, 334(6058):928--935, 2011.

\bibitem{park2016material}
Minjoon Park, Jaechan Ryu, Wei Wang, and Jaephil Cho.
\newblock Material design and engineering of next-generation flow-battery
  technologies.
\newblock {\em Nature Reviews Materials}, 2(1):1--18, 2016.

\bibitem{gurieff2019performance}
N~Gurieff, CY~Cheung, V~Timchenko, and C~Menictas.
\newblock Performance enhancing stack geometry concepts for redox flow battery
  systems with flow through electrodes.
\newblock {\em Journal of Energy Storage}, 22:219--227, 2019.

\bibitem{zhao2015chemistry}
Yu~Zhao, Yu~Ding, Yutao Li, Lele Peng, Hye~Ryung Byon, John~B Goodenough, and
  Guihua Yu.
\newblock A chemistry and material perspective on lithium redox flow batteries
  towards high-density electrical energy storage.
\newblock {\em Chemical Society Reviews}, 44(22):7968--7996, 2015.

\bibitem{hawthorne2014studies}
Krista~L Hawthorne, Jesse~S Wainright, and Robert~F Savinell.
\newblock Studies of iron-ligand complexes for an all-iron flow battery
  application.
\newblock {\em Journal of The Electrochemical Society}, 161(10):A1662, 2014.

\bibitem{petek2015slurry}
Tyler~J Petek, Nathaniel~C Hoyt, Robert~F Savinell, and Jesse~S Wainright.
\newblock Slurry electrodes for iron plating in an all-iron flow battery.
\newblock {\em Journal of Power Sources}, 294:620--626, 2015.

\bibitem{duduta2011semi}
Mihai Duduta, Bryan Ho, Vanessa~C Wood, Pimpa Limthongkul, Victor~E Brunini,
  W~Craig Carter, and Yet-Ming Chiang.
\newblock Semi-solid lithium rechargeable flow battery.
\newblock {\em Advanced Energy Materials}, 1(4):511--516, 2011.

\bibitem{backhurst1969preliminary}
JR~Backhurst, JM~Coulson, F~Goodridge, RE~Plimley, and M~Fleischmann.
\newblock A preliminary investigation of fluidized bed electrodes.
\newblock {\em Journal of the Electrochemical Society}, 116(11):1600, 1969.

\bibitem{sabacky1977electrical}
BJ~Sabacky and JW~Evans.
\newblock The electrical conductivity of fluidized bed electrodes—its
  significance and some experimental measurements.
\newblock {\em Metallurgical Transactions B}, 8(1):5--13, 1977.

\bibitem{gao2007effective}
Lei Gao, Xiaofeng Zhou, and Yulong Ding.
\newblock Effective thermal and electrical conductivity of carbon nanotube
  composites.
\newblock {\em Chemical Physics Letters}, 434(4-6):297--300, 2007.

\bibitem{dennison2014situ}
CR~Dennison, Y~Gogotsi, and EC~Kumbur.
\newblock In situ distributed diagnostics of flowable electrode systems:
  resolving spatial and temporal limitations.
\newblock {\em Physical Chemistry Chemical Physics}, 16(34):18241--18252, 2014.

\bibitem{kastening1997design}
B~Kastening, T~Boinowitz, and M~Heins.
\newblock Design of a slurry electrode reactor system.
\newblock {\em Journal of applied electrochemistry}, 27(2):147--152, 1997.

\bibitem{richards2016mixed}
Jeffrey~J Richards, Austin~D Scherbarth, Norman~J Wagner, and Paul~D Butler.
\newblock Mixed ionic/electronic conducting surface layers adsorbed on
  colloidal silica for flow battery applications.
\newblock {\em ACS applied materials \& interfaces}, 8(36):24089--24096, 2016.

\bibitem{parant2017flowing}
H~Parant, G~Muller, T~Le~Mercier, JM~Tarascon, P~Poulin, and A~Colin.
\newblock Flowing suspensions of carbon black with high electronic conductivity
  for flow applications: Comparison between carbons black and exhibition of
  specific aggregation of carbon particles.
\newblock {\em Carbon}, 119:10--20, 2017.

\bibitem{mourshed2021carbon}
Monjur Mourshed, Seyed Mohammad~Rezaei Niya, Ruchika Ojha, Gary Rosengarten,
  John Andrews, and Bahman Shabani.
\newblock Carbon-based slurry electrodes for energy storage and power supply
  systems.
\newblock {\em Energy Storage Materials}, 40:461--489, 2021.

\bibitem{li2005polymer}
Han-Ying Li, Hong-Zheng Chen, Wen-Jun Xu, Fang Yuan, Jie-Ru Wang, and Mang
  Wang.
\newblock Polymer-encapsulated hydrophilic carbon black nanoparticles free from
  aggregation.
\newblock {\em Colloids and Surfaces A: Physicochemical and Engineering
  Aspects}, 254(1-3):173--178, 2005.

\bibitem{xu2007particle}
Renliang Xu, Chifei Wu, and Haiyan Xu.
\newblock Particle size and zeta potential of carbon black in liquid media.
\newblock {\em Carbon}, 45(14):2806--2809, 2007.

\bibitem{zaccarelli2007colloidal}
Emanuela Zaccarelli.
\newblock Colloidal gels: equilibrium and non-equilibrium routes.
\newblock {\em Journal of Physics: Condensed Matter}, 19(32):323101, 2007.

\bibitem{allain1995aggregation}
C~Allain, M~Cloitre, and M~Wafra.
\newblock Aggregation and sedimentation in colloidal suspensions.
\newblock {\em Physical review letters}, 74(8):1478, 1995.

\bibitem{poon1999delayed}
W~CK Poon, LAURA Starrs, SP~Meeker, A~Moussaid, R~ML Evans, PN~Pusey, and
  MM~Robins.
\newblock Delayed sedimentation of transient gels in colloid--polymer mixtures:
  dark-field observation, rheology and dynamic light scattering studies.
\newblock {\em Faraday Discussions}, 112:143--154, 1999.

\bibitem{starrs2002collapse}
Laura Starrs, WCK Poon, DJ~Hibberd, and MM~Robins.
\newblock Collapse of transient gels in colloid-polymer mixtures.
\newblock {\em Journal of Physics: Condensed Matter}, 14(10):2485, 2002.

\bibitem{manley2005gravitational}
Suliana Manley, JM~Skotheim, L~Mahadevan, and DAVID~A Weitz.
\newblock Gravitational collapse of colloidal gels.
\newblock {\em Physical review letters}, 94(21):218302, 2005.

\bibitem{buscall2009towards}
Richard Buscall, Tahsin~H Choudhury, Malcolm~A Faers, James~W Goodwin, Paul~A
  Luckham, and Susan~J Partridge.
\newblock Towards rationalising collapse times for the delayed sedimentation of
  weakly-aggregated colloidal gels.
\newblock {\em Soft Matter}, 5(7):1345--1349, 2009.

\bibitem{teece2011ageing}
Lisa~J Teece, Malcolm~A Faers, and Paul Bartlett.
\newblock Ageing and collapse in gels with long-range attractions.
\newblock {\em Soft Matter}, 7(4):1341--1351, 2011.

\bibitem{harich2016gravitational}
Rim Harich, TW~Blythe, Michiel Hermes, Emanuela Zaccarelli, AJ~Sederman, Lynn~F
  Gladden, and Wilson~CK Poon.
\newblock Gravitational collapse of depletion-induced colloidal gels.
\newblock {\em Soft Matter}, 12(19):4300--4308, 2016.

\bibitem{ma1992mixed}
Chiming Ma and Yin Xia.
\newblock Mixed adsorption of sodium dodecyl sulfate and ethoxylated
  nonylphenols on carbon black and the stability of carbon black dispersions in
  mixed solutions of sodium dodecyl sulfate and ethoxylated nonylphenols.
\newblock {\em Colloids and surfaces}, 66(3):215--221, 1992.

\bibitem{gupta2005adsorption}
Sachin~D Gupta and Sunil~S Bhagwat.
\newblock Adsorption of surfactants on carbon black-water interface.
\newblock {\em Journal of dispersion science and technology}, 26(1):111--120,
  2005.

\bibitem{madec2015surfactant}
L{\'e}na{\"\i}c Madec, Mohamed Youssry, Manuella Cerbelaud, Patrick Soudan,
  Dominique Guyomard, and Bernard Lestriez.
\newblock Surfactant for enhanced rheological, electrical, and electrochemical
  performance of suspensions for semisolid redox flow batteries and
  supercapacitors.
\newblock {\em ChemPlusChem}, 80(2):396--401, 2015.

\bibitem{porcher2010optimizing}
W~Porcher, B~Lestriez, S~Jouanneau, and Dominique Guyomard.
\newblock Optimizing the surfactant for the aqueous processing of lifepo4
  composite electrodes.
\newblock {\em Journal of Power Sources}, 195(9):2835--2843, 2010.

\bibitem{kim2007gravitational}
Chanjoong Kim, Yaqian Liu, Angelika K{\"u}hnle, Stephan Hess, Sonja Viereck,
  Thomas Danner, L~Mahadevan, and David~A Weitz.
\newblock Gravitational stability of suspensions of attractive colloidal
  particles.
\newblock {\em Physical review letters}, 99(2):028303, 2007.

\bibitem{akuzum2017effects}
Bilen Akuzum, Lutfi Agartan, J~Locco, and EC~Kumbur.
\newblock Effects of particle dispersion and slurry preparation protocol on
  electrochemical performance of capacitive flowable electrodes.
\newblock {\em Journal of Applied Electrochemistry}, 47(3):369--380, 2017.

\bibitem{walls2003yield}
HJ~Walls, S~Brett Caines, Angelica~M Sanchez, and Saad~A Khan.
\newblock Yield stress and wall slip phenomena in colloidal silica gels.
\newblock {\em Journal of Rheology}, 47(4):847--868, 2003.

\bibitem{boehm1994some}
HP~Boehm.
\newblock Some aspects of the surface chemistry of carbon blacks and other
  carbons.
\newblock {\em Carbon}, 32(5):759--769, 1994.

\bibitem{bhatnagar2013overview}
Amit Bhatnagar, William Hogland, Marcia Marques, and Mika Sillanp{\"a}{\"a}.
\newblock An overview of the modification methods of activated carbon for its
  water treatment applications.
\newblock {\em Chemical Engineering Journal}, 219:499--511, 2013.

\bibitem{antonucci1989influence}
V~Antonucci, M~Minutoli, N~Giordano, et~al.
\newblock The influence of functional groups on the surface acid-base
  characteristics of carbon blacks.
\newblock {\em Carbon}, 27(3):337--347, 1989.

\bibitem{ridaoui2006effect}
H~Ridaoui, A~Jada, L~Vidal, and J-B Donnet.
\newblock Effect of cationic surfactant and block copolymer on carbon black
  particle surface charge and size.
\newblock {\em Colloids and Surfaces A: Physicochemical and Engineering
  Aspects}, 278(1-3):149--159, 2006.

\bibitem{medalia1964dispersant}
AI~Medalia and E~Hagopian.
\newblock Dispersant-free aqueous slurries of carbon black. viscosity,
  techniques of handling, and use in latex masterbatching.
\newblock {\em Industrial \& Engineering Chemistry Product Research and
  Development}, 3(2):120--125, 1964.

\bibitem{sis2009effect}
H{\.I}KMET Sis and MUSTAFA Birinci.
\newblock Effect of nonionic and ionic surfactants on zeta potential and
  dispersion properties of carbon black powders.
\newblock {\em Colloids and Surfaces A: Physicochemical and Engineering
  Aspects}, 341(1-3):60--67, 2009.

\bibitem{subramanian2021aqueous}
Sreedhar Subramanian and Gisle {\O}ye.
\newblock Aqueous carbon black dispersions stabilized by sodium
  lignosulfonates.
\newblock {\em Colloid and Polymer Science}, 299(7):1223--1236, 2021.

\bibitem{bossolelti1995adsorption}
Luisa Bossolelti, Riccardo Ricceri, and Glauriella Giabrielli.
\newblock The adsorption of polystyrene sulfonate and ethoxylated non-ionic
  surfactants at carbon black-water interface.
\newblock {\em Journal of dispersion science and technology}, 16(3-4):205--220,
  1995.

\bibitem{charlton2000electrolyte}
ID~Charlton and AP~Doherty.
\newblock Electrolyte-induced structural evolution of triton x-100 micelles.
\newblock {\em The Journal of Physical Chemistry B}, 104(34):8327--8332, 2000.

\bibitem{robson1977size}
Robert~J Robson and Edward~A Dennis.
\newblock The size, shape, and hydration of nonionic surfactant micelles.
  triton x-100.
\newblock {\em The Journal of Physical Chemistry}, 81(11):1075--1078, 1977.

\bibitem{bloor1970effect}
JR~Bloor, JC~Morrison, and CT~Rhodes.
\newblock Effect of ph on the micellar properties of a nonionic surfactant.
\newblock {\em Journal of pharmaceutical sciences}, 59(3):387--391, 1970.

\bibitem{paradies1980shape}
H~Hasko Paradies.
\newblock Shape and size of a nonionic surfactant micelle. triton x-100 in
  aqueous solution.
\newblock {\em The Journal of Physical Chemistry}, 84(6):599--607, 1980.

\bibitem{brown1989static}
Wyn Brown, Roger Rymden, Jan Van~Stam, Mats Almgren, and Goeran Svensk.
\newblock Static and dynamic properties of nonionic amphiphile micelles: Triton
  x-100 in aqueous solution.
\newblock {\em The Journal of Physical Chemistry}, 93(6):2512--2519, 1989.

\bibitem{garamus1998small}
Vasil~M Garamus and Jan~Skov Pedersen.
\newblock A small-angle neutron scattering study of the structure of
  graphitized carbon black aggregates in triton x-100/water solutions.
\newblock {\em Colloids and Surfaces A: Physicochemical and Engineering
  Aspects}, 132(2-3):203--212, 1998.

\bibitem{gonzalez2000determination}
CM~Gonzalez-Garcia, ML~Gonzalez-Martin, V~Gomez-Serrano, JM~Bruque, and
  L~Labajos-Broncano.
\newblock Determination of the free energy of adsorption on carbon blacks of a
  nonionic surfactant from aqueous solutions.
\newblock {\em Langmuir}, 16(8):3950--3956, 2000.

\bibitem{krall1998internal}
AH~Krall and DA~Weitz.
\newblock Internal dynamics and elasticity of fractal colloidal gels.
\newblock {\em Physical review letters}, 80(4):778, 1998.

\bibitem{lu2008gelation}
Peter~J Lu, Emanuela Zaccarelli, Fabio Ciulla, Andrew~B Schofield, Francesco
  Sciortino, and David~A Weitz.
\newblock Gelation of particles with short-range attraction.
\newblock {\em Nature}, 453(7194):499--503, 2008.

\bibitem{condre2007role}
Jean-Michel Condre, Christian Ligoure, and Luca Cipelletti.
\newblock The role of solid friction in the sedimentation of strongly
  attractive colloidal gels.
\newblock {\em Journal of Statistical Mechanics: Theory and Experiment},
  2007(02):P02010, 2007.

\bibitem{russel1991colloidal}
William~Bailey Russel, WB~Russel, Dudley~A Saville, and William~Raymond
  Schowalter.
\newblock {\em Colloidal dispersions}.
\newblock Cambridge university press, 1991.

\bibitem{buscall1987consolidation}
Richard Buscall and Lee~R White.
\newblock The consolidation of concentrated suspensions. part 1.—the theory
  of sedimentation.
\newblock {\em Journal of the Chemical Society, Faraday Transactions 1:
  Physical Chemistry in Condensed Phases}, 83(3):873--891, 1987.

\bibitem{gopalakrishnan2006linking}
V~Gopalakrishnan, Kenneth~S Schweizer, and CF~Zukoski.
\newblock Linking single particle rearrangements to delayed collapse times in
  transient depletion gels.
\newblock {\em Journal of Physics: Condensed Matter}, 18(50):11531, 2006.

\bibitem{lietor2010role}
JJ~Lietor-Santos, C~Kim, ML~Lynch, A~Fernandez-Nieves, and DA~Weitz.
\newblock The role of polymer polydispersity in phase separation and gelation
  in colloid- polymer mixtures.
\newblock {\em Langmuir}, 26(5):3174--3178, 2010.

\bibitem{bartlett2012sudden}
Paul Bartlett, Lisa~J Teece, and Malcolm~A Faers.
\newblock Sudden collapse of a colloidal gel.
\newblock {\em Physical Review E}, 85(2):021404, 2012.

\bibitem{derec2003rapid}
Caroline Derec, D~Senis, Laurence Talini, and Catherine Allain.
\newblock Rapid settling of a colloidal gel.
\newblock {\em Physical Review E}, 67(6):062401, 2003.

\bibitem{lee2006formation}
Myung~Han Lee and Eric~M Furst.
\newblock Formation and evolution of sediment layers in an aggregating
  colloidal suspension.
\newblock {\em Physical Review E}, 74(3):031401, 2006.

\bibitem{senis1997scaling}
D~Senis and C~Allain.
\newblock Scaling analysis of sediment equilibrium in aggregated colloidal
  suspensions.
\newblock {\em Physical Review E}, 55(6):7797, 1997.

\bibitem{mewis2012colloidal}
Jan Mewis and Norman~J Wagner.
\newblock {\em Colloidal suspension rheology}.
\newblock Cambridge university press, 2012.

\bibitem{masschaele2011flow}
Kasper Masschaele, Jan Fransaer, and Jan Vermant.
\newblock Flow-induced structure in colloidal gels: Direct visualization of
  model 2d suspensions.
\newblock {\em Soft Matter}, 7(17):7717--7726, 2011.

\bibitem{varga2018large}
Zsigmond Varga and James~W Swan.
\newblock Large scale anisotropies in sheared colloidal gels.
\newblock {\em Journal of Rheology}, 62(2):405--418, 2018.

\bibitem{mewis1979thixotropy}
Joannes Mewis.
\newblock Thixotropy-a general review.
\newblock {\em Journal of Non-Newtonian Fluid Mechanics}, 6(1):1--20, 1979.

\bibitem{ovarlez2013rheopexy}
Guillaume Ovarlez, Laurent Tocquer, Fran{\c{c}}ois Bertrand, and Philippe
  Coussot.
\newblock Rheopexy and tunable yield stress of carbon black suspensions.
\newblock {\em Soft Matter}, 9(23):5540--5549, 2013.

\bibitem{osuji2008shear}
Chinedum~O Osuji, Chanjoong Kim, and David~A Weitz.
\newblock Shear thickening and scaling of the elastic modulus in a fractal
  colloidal system with attractive interactions.
\newblock {\em Physical Review E}, 77(6):060402, 2008.

\bibitem{koumakis2015tuning}
Nick Koumakis, Esmaeel Moghimi, Rut Besseling, Wilson~CK Poon, John~F Brady,
  and George Petekidis.
\newblock Tuning colloidal gels by shear.
\newblock {\em Soft Matter}, 11(23):4640--4648, 2015.

\bibitem{das2021shear}
Mohan Das, Lucille Chambon, Zsigmond Varga, Maria Vamvakaki, James~W Swan, and
  George Petekidis.
\newblock Shear driven vorticity aligned flocs in a suspension of attractive
  rigid rods.
\newblock {\em Soft Matter}, 17(5):1232--1245, 2021.

\bibitem{bonacci2020contact}
Francesco Bonacci, Xavier Chateau, Eric~M Furst, Jennifer Fusier, Julie Goyon,
  and Ana{\"e}l Lema{\^\i}tre.
\newblock Contact and macroscopic ageing in colloidal suspensions.
\newblock {\em Nature Materials}, 19(7):775--780, 2020.

\bibitem{n2020yielding}
E~N’gouamba, Julie Goyon, Laurent Tocquer, Thomas Oerther, and Philippe
  Coussot.
\newblock Yielding, thixotropy, and strain stiffening of aqueous carbon black
  suspensions.
\newblock {\em Journal of Rheology}, 64(4):955--968, 2020.

\bibitem{khalkhal2018evaluating}
Fatemeh Khalkhal, Ajay~Singh Negi, James Harrison, Casey~D Stokes, David~L
  Morgan, and Chinedum~O Osuji.
\newblock Evaluating the dispersant stabilization of colloidal suspensions from
  the scaling behavior of gel rheology and adsorption measurements.
\newblock {\em Langmuir}, 34(3):1092--1099, 2018.

\bibitem{negi2009new}
Ajay~Singh Negi and Chinedum~O Osuji.
\newblock New insights on fumed colloidal rheology—shear thickening and
  vorticity-aligned structures in flocculating dispersions.
\newblock {\em Rheologica acta}, 48(8):871--881, 2009.

\bibitem{relshear}
Ryle Rel, Dennis Terwilliger, and Ryan McGorty.
\newblock Shear-induced vorticity aligned flocs in a temperature responsive
  colloid-polymer mixture.
\newblock {\em Frontiers in Physics}, page 857.

\bibitem{pool2010influence}
Ren{\'e} Pool and Peter~G Bolhuis.
\newblock The influence of micelle formation on the stability of colloid
  surfactant mixtures.
\newblock {\em Physical Chemistry Chemical Physics}, 12(44):14789--14797, 2010.

\bibitem{doi:10.1021/acs.langmuir.7b03343}
Fatemeh Khalkhal, Ajay~Singh Negi, James Harrison, Casey~D. Stokes, David~L.
  Morgan, and Chinedum~O. Osuji.
\newblock Evaluating the dispersant stabilization of colloidal suspensions from
  the scaling behavior of gel rheology and adsorption measurements.
\newblock {\em Langmuir}, 34(3):1092--1099, 2018.
\newblock PMID: 29095629.

\bibitem{hamberger2011influence}
Anika Hamberger and Katharina Landfester.
\newblock Influence of size and functionality of polymeric nanoparticles on the
  adsorption behavior of sodium dodecyl sulfate as detected by isothermal
  titration calorimetry.
\newblock {\em Colloid and Polymer Science}, 289(1):3--14, 2011.

\bibitem{yang2011electrochemical}
Zhenguo Yang, Jianlu Zhang, Michael~CW Kintner-Meyer, Xiaochuan Lu, Daiwon
  Choi, John~P Lemmon, and Jun Liu.
\newblock Electrochemical energy storage for green grid.
\newblock {\em Chemical reviews}, 111(5):3577--3613, 2011.

\bibitem{soloveichik2015flow}
Grigorii~L Soloveichik.
\newblock Flow batteries: current status and trends.
\newblock {\em Chemical reviews}, 115(20):11533--11558, 2015.

\bibitem{petek2015characterizing}
Tyler~J Petek, Nathaniel~C Hoyt, Robert~F Savinell, and Jesse~S Wainright.
\newblock Characterizing slurry electrodes using electrochemical impedance
  spectroscopy.
\newblock {\em Journal of The Electrochemical Society}, 163(1):A5001, 2015.

\bibitem{feng1984percolation}
Shechao Feng and Pabitra~N Sen.
\newblock Percolation on elastic networks: new exponent and threshold.
\newblock {\em Physical review letters}, 52(3):216, 1984.

\bibitem{gonzalez2001colloidal}
Agust{\'\i}n~E Gonz{\'a}lez.
\newblock Colloidal aggregation with sedimentation: computer simulations.
\newblock {\em Physical Review Letters}, 86(7):1243, 2001.

\bibitem{hipp2019structure}
Julie~B Hipp, Jeffrey~J Richards, and Norman~J Wagner.
\newblock Structure-property relationships of sheared carbon black suspensions
  determined by simultaneous rheological and neutron scattering measurements.
\newblock {\em Journal of Rheology}, 63(3):423--436, 2019.

\bibitem{coussot2002viscosity}
P~Coussot, Q~Dzuy Nguyen, HT~Huynh, and D~Bonn.
\newblock Viscosity bifurcation in thixotropic, yielding fluids.
\newblock {\em Journal of rheology}, 46(3):573--589, 2002.

\end{thebibliography}
		\bibliographystyle{unsrt}

\end{document}